\documentclass[preprint,authoryear,12pt,3p]{elsarticle} % 仅加宽
\usepackage[
breaklinks=true,%
pdfauthor={Peng and Bai},%
pdftitle={2017--Peng and Bai--Improving Orbit Prediction Accuracy through Supervised Machine Learning}%
]{hyperref}

\journal{Advances in Space Research}

\usepackage{graphicx} % incert figures
\usepackage{mathtools} % mathtools 取代 amsmath
\usepackage{amssymb,bm}
\usepackage{color}
\usepackage[all]{hypcap} % make the hyperlink to point at the upper border of a figure of table, rather than to point to its caption.
\usepackage{breakcites}
\usepackage{cleveref} 
\crefname{equation}{Eq.}{Eqs.} 	\Crefname{equation}{Equation}{Equations} 
\crefname{figure}{Fig.}{Figs} 	\Crefname{figure}{Figure}{Figures} 
\crefname{table}{Table}{Tables} 
\crefname{section}{Section}{Sections} % allow refer to multiple equations and change cross-reference name
\DeclareGraphicsExtensions{.pdf,.png} % to control the figures used when latex compiles
\usepackage[open]{bookmark}
\usepackage{booktabs}
\usepackage{float}
\usepackage{subfigure}
\usepackage{threeparttable} % 在 table 下面添加注释
\usepackage[section]{placeins} % 使用 \FloatBarrier 可以控制图片在subsection中
\begin{document}

%% for jasr.cls
\begin{frontmatter}
	\title{Improving Orbit Prediction Accuracy through \\ Supervised Machine Learning}

	\author{Hao Peng\fnref{footnote1}}
	\fntext[footnote1]{Postdoctoral Associate.}
	\ead{AstroH.Peng@gmail.com}
	
	\author{Xiaoli Bai\corref{cor}\fnref{footnote2}}
	\address{Mechanical and Aerospace Engineering \\ Rutgers, The State University of New Jersey, NJ 08854}
	\cortext[cor]{Corresponding author}
	\fntext[footnote2]{Assistant Professor.}
	\ead{xiaoli.bai@rutgers.edu}

	\begin{abstract}
		Due to the lack of information such as the space environment condition and resident space objects' (RSOs') body characteristics, current orbit predictions that are solely grounded on physics-based models may fail to achieve required accuracy for collision avoidance and have led to satellite collisions already. 
		This paper presents a methodology to predict RSOs' trajectories with higher accuracy than that of the current methods. 
		Inspired by the machine learning (ML) theory through which the models are learned based on large amounts of observed data and the prediction is conducted without explicitly modeling space objects and space environment, the proposed ML approach integrates physics-based orbit prediction algorithms with a learning-based process that focuses on reducing the prediction errors. 
		Using a simulation-based space catalog environment as the test bed, the paper demonstrates three types of generalization capability for the proposed ML approach: 
		1) the ML model can be used to improve the same RSO's orbit information that is not available during the learning process but shares the same time interval as the training data; 
		2) the ML model can be used to improve predictions of the same RSO at future epochs; 
		and 3) the ML model based on a RSO can be applied to other RSOs that share some common features.
	\end{abstract}
	
	\begin{keyword}
		Orbit Prediction;
		Resident Space Object;
		Supervised Machine Learning;
		Support Vector Machine.
	\end{keyword}
	
\end{frontmatter}

\parindent=0.5 cm

\section{Introduction}

%%%%%%%%%%%%%%%%%%%%%%%%%%%%%%%
% version written by Prof. Bai
%%%%%%%%%%%%%%%%%%%%%%%%%%%%%%%
The amount of resident space objects (RSOs) and the quantity of conflicts between RSOs are rapidly escalating. The number of space objects larger than 10 cm is presently approaching 21,000, the estimated population of objects between 1 and 10cm is about 500, 000, and for objects smaller than 1cm the number exceeds 100 million~\citep{nasa_orbital_2017}. 
At the heart of the challenges for Space Situational Awareness (SSA) is to predict each object's orbit efficiently and accurately. 
Several painful incidents have occurred in recent decades, such as the February 2009 collision of a U.S. Iridium communications satellite and a Russian Cosmos 2251 communication satellite~\citep{celestrak_iridium_2017}, and the RED threshold late notice conjunction threat to the International Space Station (ISS) from the ``25090 PAM-D'' debris~\citep{bergin_red_2009}. 
The limitation of the current orbit prediction capability is among the main causes for these collision events. 

Current approaches for orbit prediction are physics-based, the success of which requires a good knowledge of the state of the space object at the start of the trajectory computation, and the environment information such as earth gravity, atmospheric drag, and solar radiation pressure, as well as the intent information for the maneuvering objects. 
However, our understanding of the space/time variation of atmospheric density is limited, and our information about the space object is not accurately updated, for example the maneuvering of a spacecraft that is owned by other nations could be unavailable when the orbit prediction is conducted. 
Moreover, our current surveillance resources are limited and expensive and space-tracking measurements are sparse and noisy. 
All of these issues lead to the fact that current errors in physics-based predictions can be too large to be used for meaningful actions. 

The machine learning (ML) method presents a different modeling and prediction capability compared to the physics-based approach. 
The prediction can be made without explicitly modeling space objects, spacecraft maneuvers, and space environment. Instead, the models are learned based on large amounts of observed data, similar to human cognition in learning from past experience to predict future events. 
There are different types of machine learning, with supervised learning, reinforcement learning, and unsupervised learning as the three most common types~\citep{abu-mostafa_learning_2012}. 
Reinforcement learning is usually used to make decisions, while unsupervised learning is used to find patterns and structures within the data such as to cluster data into different groups without providing outputs to describe the groups. 
Supervised learning is such a method that learns a function or mapping from labeled data. 
The training data are pairs of input and its corresponding output. 
The supervised ML method is the appropriate approach for improving orbit prediction error based on historical measurements and error information. 
In the remainder of this paper, for brevity, we will refer to supervised machine learning methods as ML methods.

The ML methods have shown great capability for a wide range of applications~\citep{abu-mostafa_learning_2012,mitchell_machine_1997}, including many in the aerospace field~\citep{ampatzis_machine_2009,hartikainen_state-space_2012,sharma_robust_2015-2}. 
Hartikainen et al. combine physical models with non-parametric data-driven techniques to build a model for orbit prediction~\citep{hartikainen_state-space_2012}. Their method focuses on the data mining aspect and can extract information of unknown forces from historical data. 
Sharma and Cutler have presented a learning approach to do orbit determination based on distribution regression and transfer learning methods~\citep{sharma_robust_2015-2}. Their tests show that the proposed machine learning approach is superior to the conventional methods such as the extended Kalman filter. Moreover, the method is able to estimate significantly varying orbital parameters. 

This paper presents a computational framework that improves orbit prediction accuracy through the ML method. A crucial feature of the proposed approach is that the overall framework will introduce and embed a learning-centered strategy into the physics-based prediction. To take advantage of the knowledge on the physics-based models that are important and represent the state of the practice, we design the learning process to focus on modeling the prediction errors instead of the full dynamics. 
In this way, the prediction errors introduced in different steps including measurement, estimation, and prediction, are captured as a surrogate model. % under the machine learning approaches. 
An extra benefit from this approach is that the learning is only burdened with finding the incremental corrections to the physics-based prediction, which reduces the dimensionality for the learning task. The support vector machine (SVM) regression method~\citep{smola_tutorial_2004}, which is an advanced method proven to be robust to outliers in the dataset, is chosen in this paper, although the proposed methodology is expected to be general and applicable to other ML methods. Furthermore, in contrast with other researches that focus on TLE data~\citep{legendre_two_2006,sang_experimental_2014,sang_estimation_2013,wang_propagation_2009}, we use a simulation-based space catalog environment to validate the proposed orbit prediction method.  This approach is important, because a known ``truth'' will underlie all simulations, and the solutions from the machine learning can be compared to the ``truth'' to make definitive statements on the performance.  

Our simulation results, interestingly and innovatively, demonstrate three types of generalization capability for the proposed ML approach. 
First, the ML model can be used to improve the RSO's orbit information that is not available during the learning process but share the same time interval as the training data. 
Second, the ML model can be used to improve predictions of the same RSO at future epochs. 
And third, the ML model based on one RSO can be applied to other RSOs that share some common features. 
Our results also provide insight upon several questions that are important for the SSA applications including:
more observation stations could generate more historical data and thus better performance can be expected from the ML model; 
larger size of the training data can lead to better performance but there appears to be a limit; 
and the generalization capability to nearby RSOs implies that it is possible to build an ML system for a large number of RSOs while only representative RSOs are used for training. 

The paper outline is as follows. The high-fidelity simulation environment is first described in \cref{sec:Simulation Environment}. 
Next, the framework of the machine learning approach to improve orbit prediction accuracy is presented in \cref{sec:Machine Learning}. 
In \cref{sec:numerical results}, the numerical simulation results are analyzed, where a performance metric is proposed to evaluate the ML approach. 
In \cref{sec:generalize to future}, the generalization problem to future epochs is proposed and evaluated for SSA application. 
Then in \cref{sec:generalize to other RSOs}, the generalization problem to different RSOs is examined by varying the inclination and the semi-major axis of the simulated RSO. 
Finally, conclusions and future studies are presented in the last section.

\section{Simulation Environment}
\label{sec:Simulation Environment}

\begin{figure}[!htbp]
	\centering
	\includegraphics[width=0.8\linewidth]{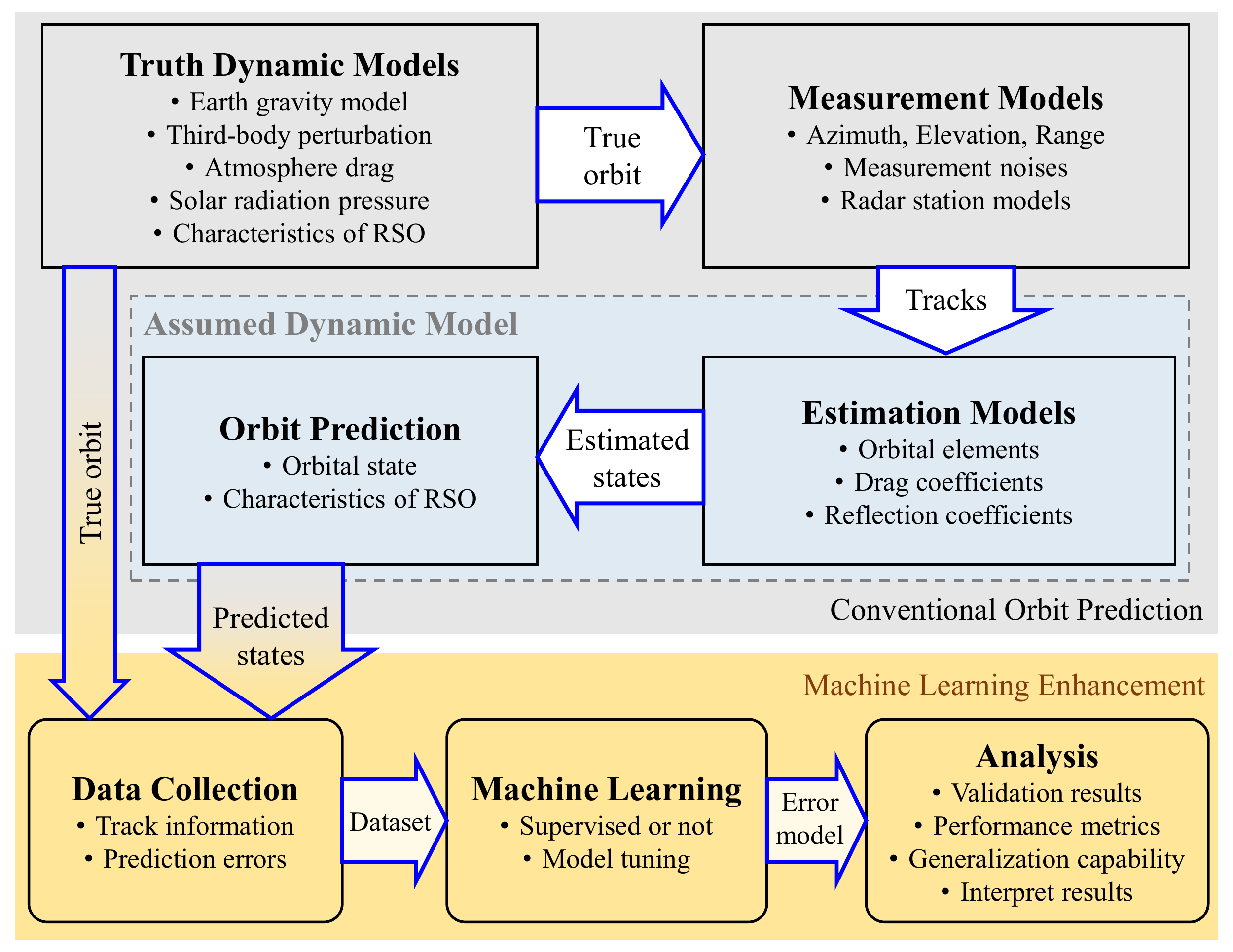}
	\caption{The framework of improving orbit predictions through the ML method.}
	\label{fig:flow_chart}
\end{figure}

The framework of the orbit predictions through supervised machine learning is shown in \cref{fig:flow_chart}. 
The top four blocks show the conventional orbit prediction process, while the bottom three blocks represent the proposed ML approach to enhance the conventional orbit prediction. 
In the following subsections, the top four blocks will be explained, and the bottom three blocks will be presented in \cref{sec:Machine Learning}. 

\subsection{Truth Dynamic Models}

The truth dynamic models include
\begin{itemize}
	\item Basic Newtonian two-body gravitational force, with an accurate gravitational constant.
	\item High-fidelity non-spherical gravity model of the Earth.
	\item Third-body perturbations.
	\item Air drag force model with high-fidelity atmosphere model.
	\item Solar radiation pressure model with high-fidelity solar activity model.
%	\item Solid tides, ocean tides, etc.
	\item Other parameters depending on the specific RSO, including the mass, inertia, geometry, material properties, etc.
\end{itemize}
The choice of truth dynamic model for the simulation depends on the time scale. 
In this paper, the truth dynamic model is expected to include the major factors that could contribute to the orbit prediction error within at least 60 days. 
Therefore, forces that have long-period effects do not need to be modeled with high precision. 
The setup of the truth model is summarized in \cref{tab:true and assumed models}. 
The Newtonian gravitational force is added with an Earth gravitational constant of $3.986\times10^{14} {\rm m}^3/{\rm s}^2$. 
The non-spherical effect of the Earth gravity is modeled using spherical harmonic functions, with coefficients provided by the EIGEN-6S model~\citep{forste_new_2012}, truncated with degree/order $40 \times 40$ as the truth gravity. 
%The EIGEN-6S is a very accurate satellite-only gravity field of a maximum degree/order 240, consisting of 6.5 years of LAGEOS (SLR) and GRACE (GPS-SST and K-band range rate) data from the time span 1 Jan 2003 till 30 June 2009 and 6.7 months of GOCE (Satellite gradiometry) data from the time span 1 Nov 2009 till 30 June 2010. 
%
Third-body perturbations of all major solar system bodies are considered, including the Sun, all the planets, the Pluto, and the Moon. 
The position of these bodies are provided by DE430 data file from the JPL~\citep{folkner_planetary_2014}. 
%
%The atmosphere cannot be precisely modeled and the exponential model is often used to calculated the drag force~\citep{lubey_identifying_2014}. 
%To provide a more realistic truth for our study, we choose to use 
The DTM2000 model is used to approximate the atmosphere. 
The Marshall Solar Activity Future Estimate Solar (MSAFE) data from NASA is used to provide solar activity information, which has significant effect on the density and the speed of the atmosphere. 
The solar radiation pressure is calculated with the reference value as $4.56\times 10^{-6}$ $\rm N/m^2$ at 1 AU (149,597,870.0 km) from the Sun. And the effect of the penumbra and eclipse are considered. 
%
%Since no particular kind of RSO is specified 
Additionally, a spherical RSO with a constant section area is assumed, and the drag coefficient $C_d$ and single-parameter reflection coefficient $C_r$ are assumed to be constant. 
These models are implemented using the Orekit, which is a low level space dynamics library written in Java~\citep{maisonobe_orekit:_2010}. %, as well as the models in the following subsections. 
%It provides a validated platform where we have complete controls over our programs and simulations, such as incorporating new conditions and constraints for the simulation. 

\begin{table*}[!htbp]
	\centering
	\begin{threeparttable}
		\caption{Parameters of the ``truth'' model used to generate orbits and measurements, and the assumed model used in the estimation and prediction.}
		\label{tab:true and assumed models}
		\footnotesize
		\begin{tabular}{rll}
			\toprule
			             Parameters & Truth model         & Assumed model \\ \midrule
			            Earth Shape & WGS84               & WGS84         \\
			Harmonics Gravity Field & $40\times40$        & $20\times20$  \\
			Third-Body Purterbation & Sun + Solar Planets & Sun + Jupiter \\
			                        & + Pluto + the Moon & + the Moon    \\
			      Atomosphere Model & DTM2000             & DTM2000       \\
			   Solar Activity Model & MSAFE               & MSAFE         \\ \bottomrule
		\end{tabular}
	\end{threeparttable}
\end{table*}

\subsection{Measurement Models}

Measurements of a RSO can be obtained using ground-based radar or optical stations~\citep{hill_comparison_2012}, or space-based systems using specially designed satellite~\citep{lyon_geosynchronous_2004-1}. 
In this study, only ground radar stations are considered because the paper focuses on the general machine learning framework and the target RSO is assumed to be an LEO object. 
In the simulations, the radar station is located at a topocentric frame centered at a given geodetic point location defined on the WGS84 Earth ellipsoid. 
At each step of orbit propagation, the visibility of the RSO with respect to the ground stations is checked. 
If the RSO is visible to a station, the station will provide measurements including the azimuth $\alpha$, the elevation $\eta$, and the range $\rho$. 
These measurements are generated by converting the state vector of the RSO in the Earth Centered Inertial (ECI) frame~\citep{vallado_fundamentals_1997} to the topocentric frame and then calculating the angle information. 
A series of consecutive measurements are organized as a track. 
%The track of a RSO is defined as a continuous arc along its true orbit. 
If more than one station detect the RSO at a particular time, all the stations will generate measurements which are combined into one track. %, regardless of the detail that how their data are shared.
Therefore, one track will start when it is visible to any station, and end when no stations can detect it.

% \usepackage{booktabs}
\begin{table*}[!htbp]
	\centering
	\caption{Ground-based radar stations modeled in the paper~\citep{hill_comparison_2012}.}
	\label{tab:stations}
	\begin{tabular}{llll}
		\toprule
		                 Station                   & Eglin, FL & Clear, AK & Kaena Point, HI \\ \midrule
		         Latitude [deg]           &   30.57   &   64.29   &      21.57      \\
		         Longitude [deg]          &  $-$86.21   &  $-$149.19  &     $-$158.27     \\
		             Altitude {[}m{]}              &   34.7    &   213.3   &      300.2      \\
		       Maximum Range $\rho$ {[}m{]}        &   13210   &   4910    &      6380       \\
		Feasible Elevation $\alpha$ [deg] &   1--90   &   1--90   &      0--90      \\
		          $\sigma_\rho$ {[}m{]}            &   32.1    &   62.5    &      92.5       \\
		      $\sigma_\alpha$ [deg]       &  0.0154   &  0.0791   &     0.0224      \\
		       $\sigma_\eta$ [deg]        &  0.0147   &   0.024   &     0.0139      \\ \bottomrule
	\end{tabular}
\end{table*}

In the simulations, three radar stations are simulated, with their parameters listed in \cref{tab:stations}. 
The RSO is visible to a station only if the range is less than the maximum range and the elevation is within the feasible elevation range. 
%A constant observation gap is assumed. 
%When the RSO is visible, the true observations without errors are directly recorded in data files, tagged with the station from with it is generated. 
%Then the measurement errors are added accordingly when reading observation data later. 
%Handling data in this way provides a convenience way to test different measurement errors without re-propagate the true orbit. 
Measurements are given at a 20-second interval. 
The measurement errors are simulated as normal distributions with zero biases and standard deviations of $\sigma_\alpha$, $\sigma_\eta$, and $\sigma_\rho$ for the azimuth, elevation and range respectively, as listed in \cref{tab:stations}. 
We remark that the constraint of the azimuth range of the station~\citep{hill_comparison_2012} has been omitted in the current study, so the error model does not depends on the azimuth.

\subsection{Estimation Models}

\begin{figure*}[!htbp]
	\centering
	\includegraphics[width=0.9\linewidth]{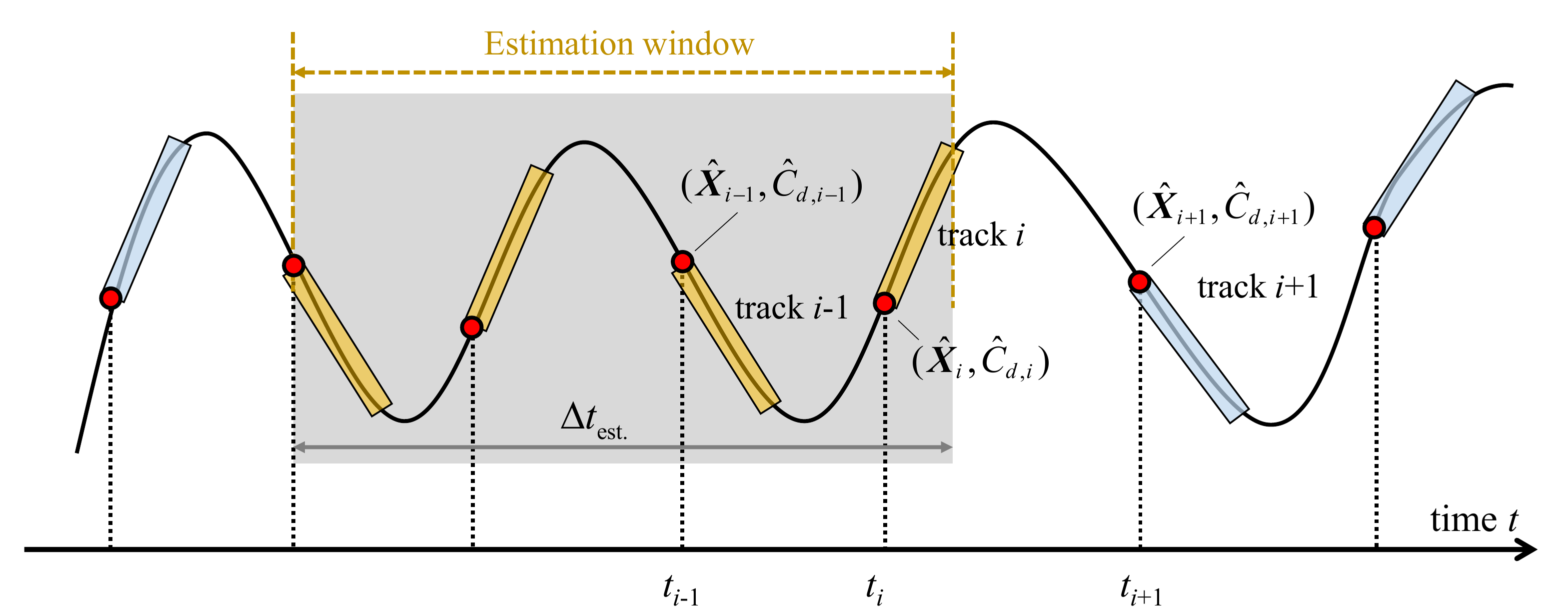}
	\caption{Illustration of the tracks included for an estimation at $t_i$, providing the estimated state $\hat{\bm X}_i$ and drag coefficient $\hat{C}_d$ of the RSO.}
	\label{fig:tracks_for_estimation}
\end{figure*}

\Cref{fig:tracks_for_estimation} illustrates the estimation strategy. 
The black curve represents the true orbit of a RSO.
The colored (light orange and blue) boxes along the curve represent observation tracks. 
The estimation is carried out at the beginning epoch of each track, as represented by the red dots in the figure. 
For the $i$-th track in the figure, whose beginning epoch is $t_i$, measurements on current track and all previous tracks $t_b$ with $t_b-t_i\leq\Delta t_\text{est.}$ are considered, where $\Delta t_\text{est.}=12$ hours is the chosen estimation window. 
This estimation window is shown by the gray rectangle underlying the true orbit in \cref{fig:tracks_for_estimation}.
Orange rectangles show all tracks in the estimation window that will be used to produce the estimation $(\hat{\bm X}_i,\hat{C}_{d,i})$ at $t_i$, while blue rectangles show tracks outside the estimation window.
We remark that this strategy is adopted to simulate the realistic process of the orbital determination, where single observation track may not have enough measurements. 

The batch Least Square (LS) estimator~\citep{crassidis_optimal_2011} is chosen to estimate the state of the RSO. 
% revision part
In practice, the LS estimator is only sub-optimal due to the nonlinearity of the model. 
However, the focus of this paper is the framework of the proposed ML approach and our hope is that it can work with most of the specific estimation methods. 
Additionally, we note that although the batch least square estimator can also generate the covariance matrix, which is important for collision risk assessment, only the orbital state is used in our simulation. 
This is due to the fact that the covariance information is not available for the two-line element (TLE) catalog~\citep{vallado_revisiting_2006}. 

Starting from the block of estimation models in \cref{fig:flow_chart}, simulations are carried out using an assumed dynamic model. 
In practice, the assumed dynamic model used during estimation and propagation is often only an approximation to the real physics. 

In the simulations, the assumed model is set up with the spherical harmonic gravity model of degree/order $20 \times 20$, and with the third-body perturbations including only three major resources, the Sun, the Moon and the Jupiter, as summarized in \cref{tab:true and assumed models}. 
The orbit state $\bm{X}$ and the drag coefficient $C_d$ of the RSO are estimated by the LS estimator, which generates the pairs of estimations denoted as $(\hat{\bm{X}},\hat{C}_d)$. 
Other parameters of the RSO are fixed, including the cross section area, the mass, and the reflection coefficient $C_r$ of the surface. 
Finally all pairs of estimations $(\hat{\bm X},\hat{C}_{d})$ are recorded for the orbit predictions. 
We remark that if the differences between the true model and the assumed model are smaller, the error of the estimation will be smaller, and also the deviations of propagating two estimates to the same epoch will be smaller. 
As a consequence, compared with other assumed models with larger differences, less information is embedded in the simulated catalog using the current assumed model. 
Using such conservative assumptions, the results in \cref{sec:numerical results,sec:generalize to future,sec:generalize to other RSOs} show that the ML approach improves the orbit prediction accuracy. 
Therefore, we demonstrate the capability of the ML approach in a more significant way.

\subsection{Orbit Prediction}
\label{subsec:simulation of predictions}
After obtaining estimations $(\hat{\bm X}_i,\hat{C}_{d,i})$ for all the tracks, the prediction process is straightforward. 
Using the assumed dynamic model, which is the same as the one that has been used for estimation, as show in \cref{fig:flow_chart}, each estimated state $\hat{\bm X}_i$ is propagated to a future epoch $t_j$ with $t_j > t_i$, when another observation track begins. 
%By doing this, we assumed that orbit prediction is used to provide a guess of the next transit of the RSO for a continuous space surveillance observation. 
At the future epoch $t_j$, the predicted state $\hat{\bm{X}}_{j;i}$ based on $\hat{\bm{X}}_i$ is compared with the true state $\bm X_j$, which provides the prediction errors. 
%All relevant variables and information should be collected in this process for the machine learning methods. 
Since the assumed dynamic model, the measurement process, and the estimation process will all introduce errors, the resulting prediction error can grow quickly to a meaningless magnitude as the propagation time increases. 
Therefore, we use a maximum prediction duration $\Delta t_{\rm max}=7$ days to restrict the orbit prediction process in this paper. 
This duration is also long enough for the surveillance and collision avoidance for the LEO objects. 

%It is usually a trail-and-error process to tune the performance of the ML method, and the setup of above simulation blocks allow us to manipulate each blocks without interrupting its foregoing blocks. 

\section{Prediction Accuracy Improvement through Machine Learning}
\label{sec:Machine Learning}

In this section, the proposed ML approach to improve orbit prediction accuracy is presented. 
We first introduce the SVM model, which is the adopted machine learning method in this paper. 
We then present the machine learning model of the orbit prediction error. 
And we discuss the dataset chosen for the SVM model to capture the orbit prediction error in the last subsection.

\subsection{Support Vector Machine}

The support vector machine (SVM) method is a machine learning algorithm that can be used for both classification and regression problems. 
One strength of the SVM method is that they are nonparametric techniques, so we do not need to specify the basis functions in prior. 
The SVM regression can handle nonlinear problems since it relies on kernel functions. 
Moreover, the SVM method has universal approximation capability with various kernels including the Gaussian kernel~\citep{hammer_note_2003,micchelli_universal_2006}. 
We note that although it cannot be verified that the orbit prediction problem under this study satisfies the fundamental assumptions of SVM's universal approximation property, the result in this paper demonstrate that the SVM method is practically effective to reduce orbit prediction errors. 
The concept of the SVM is briefly reviewed below, and the details are referred to the references.

The $\varepsilon$-SVM regression method aims to find a function $f(x)$ that has at most $\varepsilon$ deviation from the actual obtained targets for all the training data, and at the same time $f(x)$ is required to be as flat as possible~\citep{smola_tutorial_2004}. 
Suppose we have a set of training data $\mathcal{X} = \{ (\bm{x}_1,y_1),\allowbreak (\bm{x}_2,y_2),\allowbreak \dots,\allowbreak (\bm{x}_n,y_n) \} \in \mathcal{L}\times\mathbb{R}$, where $\mathcal{L}$ denotes the space of input learning variables. 
In the linear case, the desired function takes the form 
\begin{equation} \label{eq:f(x)}
	f(\bm{x}) = \langle \bm{w},\bm{x} \rangle + b,
\end{equation}
where $\bm{\omega}\in\mathcal{L}$, $b\in\mathbb{R}$, and $\langle \cdot,\cdot \rangle$ represents the inner product operation. 
Then the training problem is to find the flattest function in the space $\mathcal{L}$, where the flatness of the function is represented by $ \| \bm{\omega} \|^2 $. 
The training problem is cast as a convex optimization problem to minimize the cost function
%\begin{equation}
	$ \frac{1}{2} \| \bm{\omega} \|^2 $,
%\end{equation}
subject to the constraint 
%\begin{equation}
	$ | y_i - \langle \bm{\omega}, \bm{x}_i \rangle - b | \leq \varepsilon $
%\end{equation}
for all the training data. 
By introducing slack variables and then solving the dual problem of the above quadratic problem, the support vector expansion of the regression function $f(x)$ can be expresses as
\begin{equation} \label{eq:f(x) sv expansion}
	f(\bm{x}) = \sum_{i}^{} (\alpha_i-\alpha_i^*) \langle \bm{x}_i, \bm{x} \rangle + b,
\end{equation}
where $\alpha_i$ and $\alpha_i^*$ are the dual variables solved from the dual problem~\citep{smola_tutorial_2004}. 
Comparing \cref{eq:f(x),eq:f(x) sv expansion}, we can see that $\bm{\omega}$ can be completely described as a linear combination of the training data $x_i$. 
This property of the SVM makes it possible to deal with nonlinear regressions via kernels. 
Substituting the inner product $\langle \cdot,\cdot \rangle$ in $\cref{eq:f(x) sv expansion}$ by the kernel $k(\cdot,\cdot)$, the optimization problem is reformulated to find a flattest function in the feature space indicated by the kernel, expressed as
\begin{equation} \label{eq:f(x) kernel}
	f(\bm{x}) = \sum_{i}^{} (\alpha_i-\alpha_i^*) k(\bm{x}_i,\bm{x}) + b.
\end{equation}

As shown in \cref{eq:f(x) kernel}, when using the kernel technique, the coefficient $\bm{\omega}$ in \cref{eq:f(x)} will not be provided explicitly. 
However, when given a new testing data ${\bm{x}}_{\rm test}$, according to \cref{eq:f(x) kernel}, we only need the kernel $k(\cdot,\cdot)$ and the corresponding dual variables $\alpha_i$ and $\alpha_i^*$ to find the prediction $f(\bm{x}_{\rm test})$ by the trained SVM regression model.

\subsection{Machine Learning Model of the Prediction Error}

\begin{figure}[!htbp]
	\centering
	\includegraphics[width=0.48\textwidth]{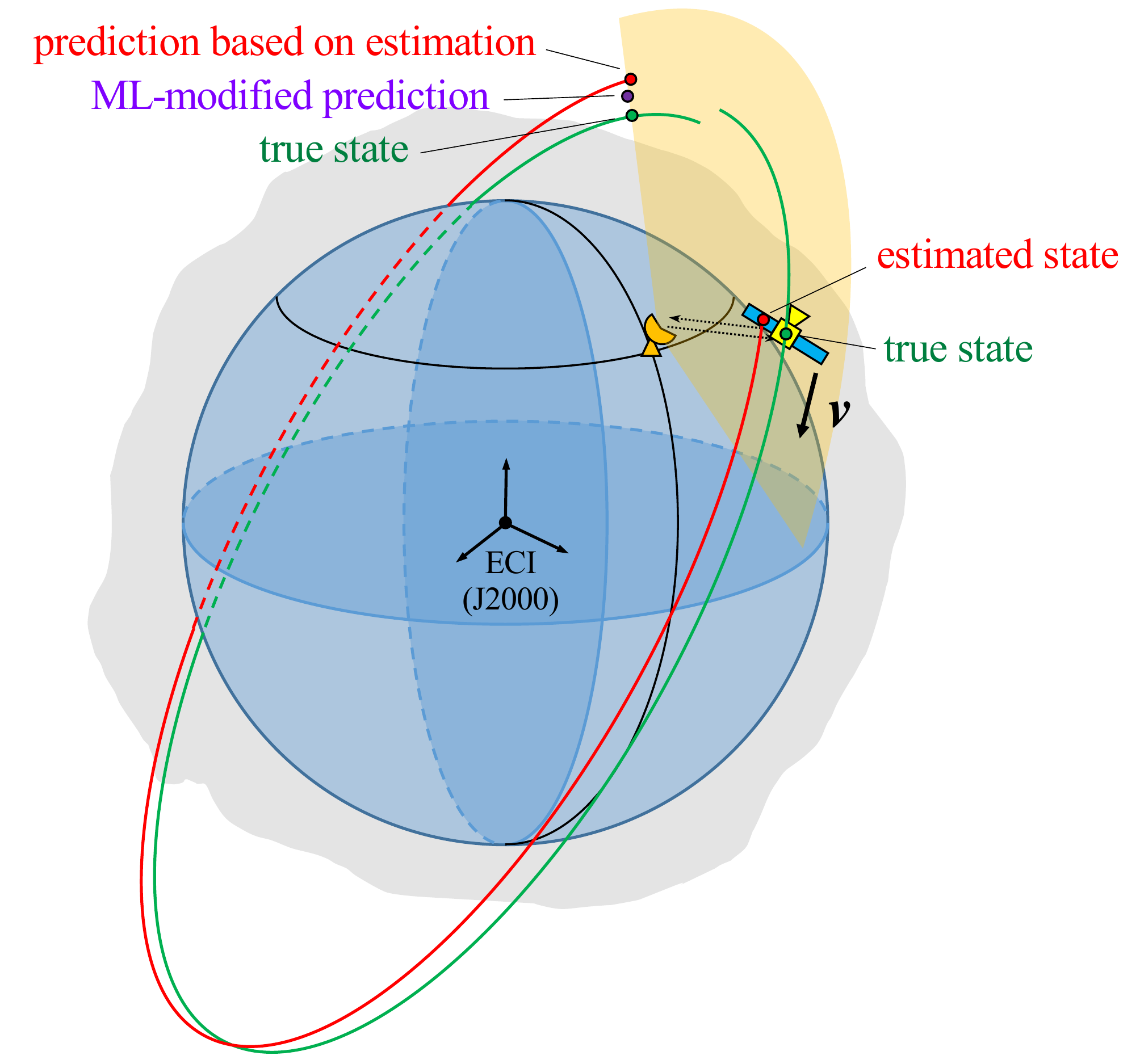}
	\caption{Modify the prediction through ML-based correction.}
	\label{fig:prediction_illustration}
\end{figure}

The concept of the ML approach to modify the prediction is illustrated in \cref{fig:prediction_illustration}. 
After the ground station observes a RSO, the estimation of the RSO is conducted. 
The estimated state will have errors, and the following orbit prediction will further deviate from the true orbit because of the assumed dynamic models. 
We introduce the ML approach to directly modify this prediction, such that the ML-modified prediction will be closer to the true state. 
Importantly, as shown in \cref{fig:prediction_illustration}, this modification does not address the state around the estimation at the current epoch, while in contrast, it directly improves the prediction at the future epoch. 

\FloatBarrier
\subsection{Dataset Capturing the Orbit Prediction Error Model}

The structure of the training data should be carefully designed to capture the prediction errors. 
%From theoretical point of view, the more variables are collected, the more information about the system could be learned. 
%However, in practice, not all variables can be obtained. 
%Usually we have only limited measure capability on certain variables, and even this limited measure capability is with noise. 
We note that although the dataset designed in this paper is not the unique solution, results in the paper show that the ML approach can indeed improve the orbit prediction accuracy based on the proposed design of the dataset. 
%Moreover, different structures of the dataset can lead to different performances, and this is still an ongoing research. 

\begin{figure*}[!htbp]
	\centering
	\includegraphics[width=0.95\textwidth]{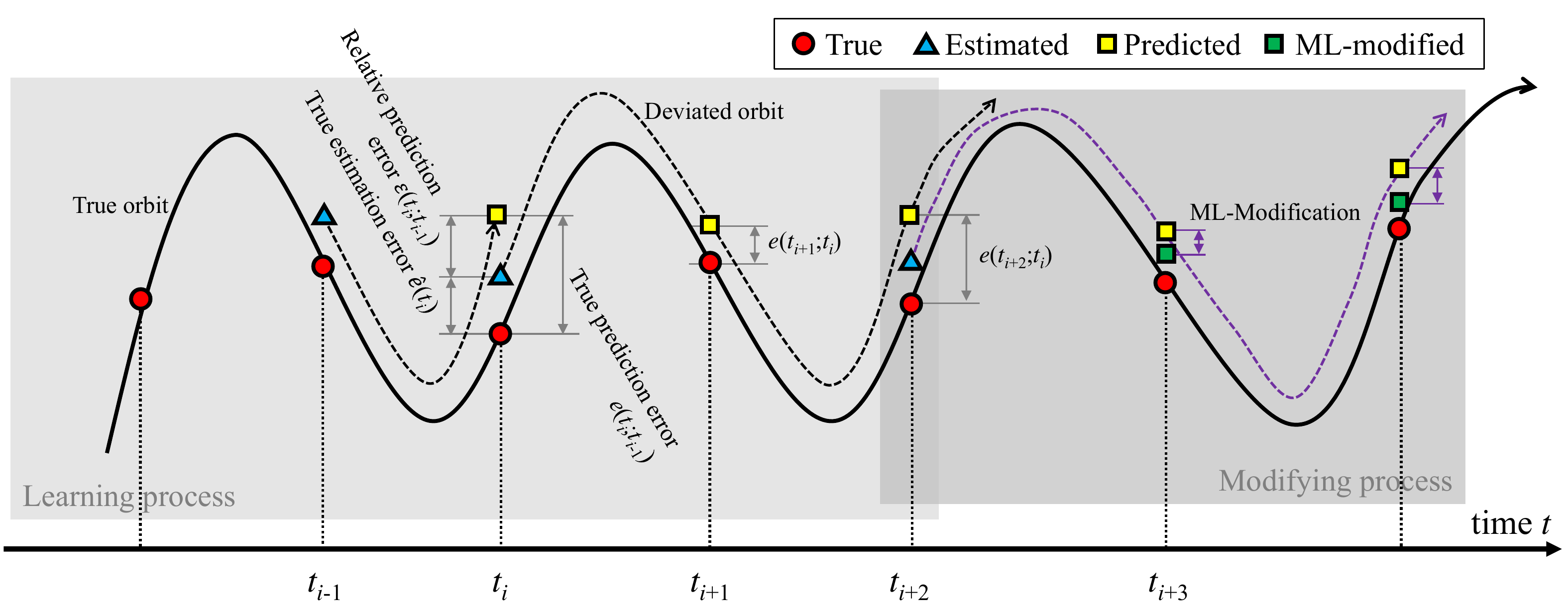}
	\caption{Illustration of the dataset for the ML approach to improve orbit prediction accuracy.}
	\label{fig:learning_dataset}
\end{figure*}

\Cref{fig:learning_dataset} illustrates the construction of the dataset for the ML approach in this study. 
The horizontal axis represents the time $t$. 
The dark solid curve represents the true orbit, and the dashed curves represent the predicted orbit arcs propagated based on the assumed dynamic models. 
%The true state of the RSO in the inertial frame at the epoch $t_j$ is denoted as $\bm{X}_j = \left.[x,y,z,v_x,v_y,v_z]^{\rm T}\right|_{t_j}\in\mathbb{R}^6$. 
%Different markers represent specific state of the RSO, including the true state $\bm{X}_j$, estimated state $\hat{\bm{X}_j}$, predicted state $\hat{\bm{X}}_{j;i}$ (based on $\hat{\bm{X}}_j$, and ML-modified states $\hat{\bm X}_{{\rm ML},j;i}$, as shown by the legend in the figure.
The true state of the RSO at any specified epoch $t$ is denoted as $\bm{X}(t)$, which can be expressed in different coordinate frames and forms. 
In the classical orbital elements (COE) form, the state $\bm{X}(t)$ is expressed as ${}^{\mathsf{COE}}\bm{\gamma}(t)=[a,e,i,\omega,\Omega,f]^{\rm T}\in\mathbb{R}^6$. 
In the Cartesian coordinates of the Earth-centered inertial (ECI) frame, the state $\bm{X}(t)$ is expressed as ${}^{\mathsf{ECI}}\bm{x}(t) = [X,Y,Z,V_X,V_Y,V_Z]^{\rm T}\in\mathbb{R}^6$.
%Different markers in the figure represent specific states of the RSO. 
At a specific time $t_i$, we have the true state $\bm{X}(t_i)$, estimated state $\hat{\bm{X}}(t_i)$, predicted state $\hat{\bm{X}}(t_i;t_b)$, and ML-modified state $\hat{\bm{X}}_{\text{ML-modified}}(t_i;t_b)$, where the second time parameter $t_b$ in the parenthesis indicates the prediction is based on the previous estimation at an earlier epoch $t_b\leq t_i$.

In \cref{fig:learning_dataset}, the lighter gray box on the left hand side shows the learning process of the ML method, during which the training data are collected to train the SVM model. 
As annotated in the figure, at the epoch $t_{i}$, there are three states about the RSO, the true state $\bm{X}(t_{i})$, the estimated state $\hat{\bm{X}}(t_{i})$, and the predicted state $\hat{\bm{X}}(t_{i};t_{i-1})$ which is based on the previous estimation $\hat{\bm{X}}_{i-1}$. 
We have three types of error (as shown in \cref{fig:learning_dataset}): 
\begin{itemize}
	\item True prediction error, $\bm{e} (t_{i};t_{i-1}) = \hat{\bm{X}}(t_i;t_{i-1}) - \bm{X}(t_{i})$. 
	
	\item True estimation error, $\hat{\bm{e}} (t_{i}) = \hat{\bm{X}}(t_{i}) - \bm{X}(t_{i})$.
	
	\item Relative error of prediction, $\bm{\varepsilon} (t_{i};t_{i-1}) = \hat{\bm{X}}(t_{i};t_{i-1}) - \hat{\bm{X}}(t_{i})$. 
\end{itemize}
These three errors are related by the equation 
\begin{equation}
	\bm{e} (t_{i};t_{i-1}) = \hat{\bm{e}} (t_{i}) + \bm{\varepsilon} (t_{i};t_{i-1}). 
\end{equation} 
Notice that the relative prediction error $\bm{\varepsilon} (t_{i};t_{i-1})$ is identical to the definition of the consistency error between two estimations in the work of \citet{rivera_improving_2016}. 
These orbit prediction errors often represented in the RSW coordinate frame, where $x$-axis (radial) is the radial direction, the $y$-axis (along-track) is perpendicular to the $x$-axis in the orbital plane and points to the inertial velocity direction, and the $z$ axis (cross-track) is along the angular momentum direction~\citep{vallado_fundamentals_1997}.

Next, we construct the dataset for the ML model. 
Apart from the three types of errors, there are other variables available at $t_{i}$, such as the measurement of the current track, which could also contain information about the orbit prediction error. 
Take the epoch $t_{i}$ as an instance, and assume the previous estimation is obtained at the epoch $t_b$, with $t_b\leq t_i$. 
Then, the learning variables of a particular data point of the dataset consist of the following: 
\begin{itemize}
	\item The relative prediction error $\bm{\varepsilon}(t_i;t_b)$ based on the previous estimation $\hat{\bm{X}}(t_{b})$, expressed in both the form of COE as ${}^{\mathsf{COE}}\delta \bm \gamma_{\bm{\varepsilon}}$, and the form of RSW frame as ${}^{\mathsf{RSW}}\delta {\bm x}_{\bm{\varepsilon}}$. 
	So we will have $\bm{\Delta} = [{}^{\mathsf{COE}}\delta \bm{\gamma}_{\bm{\varepsilon}}^{\rm T}, {}^{\mathsf{RSW}}\delta {\bm x}_{\bm{\varepsilon}}^{\rm T}]^{\rm T}$ as a representation of $\bm{\varepsilon}(t_i;t_b)$, with redundancies. 
	It is expected that $\bm{\Delta}$ carries the information of the model error between the assumed model and the truth model. 
	
	\item The prediction duration $\Delta t = t_{i}-t_{b}$. 
	This information is an important factor, as the longer the prediction, the larger the prediction error. 
	
	\item Current estimation of the orbit state, expressed in the COE form as ${}^{\mathsf{COE}} \hat {\bm\gamma}(t_i) $. 
	This information can reflect physical information of the orbit, such as the altitude and the shape. The atmosphere drag force and the solar radiation force also depend on these information. 
	
	\item The estimated drag coefficient $\hat{C}_d(t_{b})$ at the previous estimation, and its error $\delta \hat{C}_d(t_i;t_b)$ ${}={}$$\hat{C}_d(t_{b})$${} - \hat{C}_d(t_{i})$ with respect to the current estimation $\hat{C}_d(t_{i})$. 
	This information is expected to be important for RSOs in LEO. 
	
	\item Maximum elevation $\eta$ at the current track starting with $t_{i}$, and the corresponding azimuth $\alpha$ and range $\rho$. 
	This information is usually located around the middle of the track. They can reflect the difficulty and accuracy of the measurements along the track, because when the RSO has a larger range and smaller elevation, the sensitivity of the estimated state to the measurement will be smaller. 
	
\end{itemize} 
We note that it is not necessary to distinguish the estimation error and the propagation error for the proposed ML approach. 
The target variable for the above data point is chosen as:
\begin{itemize}
	\item True prediction error $\bm{e}$ at a future epoch $t_j$ ($t_i < t_j$), expressed in the RSW frame as ${}^{\mathsf{RSW}}\delta \bm{x}_{\bm{e}}=[e_{x},\allowbreak e_{y},\allowbreak e_{z},\allowbreak e_{vx},\allowbreak e_{vy},\allowbreak e_{vz}]^{\rm T}\in\mathbb{R}^6$. 
\end{itemize}
The illustrated $\bm{e}(t_{i+1};t_i)$ and $\bm{e}(t_{i+2};t_i)$ in \cref{fig:learning_dataset} are examples of the target variable, based on the same estimation $\hat{\bm{X}}_{i}$ but with different prediction durations of $\Delta t = t_{i+1}-t_{i}$ and $t_{i+2}-t_{i}$ respectively. 
With $^{\mathsf{RSW}}\delta \bm{x}_{\bm{e}}$ being a 6-dimensional vector, six SVM models in total will be trained for all the components. 

With these pairs of learning and target variables, the ML model is designed to directly modify a future orbit prediction.
As shown by the prediction process in \cref{fig:learning_dataset} (the darker gray block at the right side), the predicted state is adjusted by the additional ML-modification, which is the output of the ML model. 
If the learning and target variables are related, and if their relationship is at least partially contained in the designed dataset, the ML model should capture the underlying relationship and the ML-modification is expected to bring the prediction closer to the true orbit.

%The learning variables $\bm\Lambda$ at $t_i$ is collected as input to the SVM model. 
%Then we can modify the prediction at the future epoch $t_j$ ($j>i$) by subtracting the ML-predicted true error $\hat{\bm{e}}_{\rm ML}$, from the prediction $\hat{\bm{X}_{i;j}$ based on the previous estimation $\hat{\bm X}_i$. 

Collect all learning variables as $\bm \Lambda = [\Delta t, \bm\Delta, \hat{\bm\gamma}, \alpha, \eta, \rho]$. 
One data point of the dataset is expressed as $\left(\bm\Lambda(t_i), e_{\xi}(t_j)\right)$, where $\xi\in\{x,y,z,vx,vy,vz\}$ corresponds to different state components of the target true error $\bm{e}$, and $t_i$, $t_j$ ($t_i < t_j$) indicate the different epochs of the learning variables $\bm{\Lambda}$ and the target variable $e_{\xi}$ respectively. 
Then finally, the whole dataset $\mathcal{X}$ is constructed as, 

\newcommand{\phterm}[2]{\left(\bm\Lambda(t_{#1}), e_{\xi}(t_{#2})\right)}
\begin{align*}
	\mathcal{X}  =	
				\{ & \dots &  & \dots              &  & \dots              &  & \dots            &  & \dots          \\
	               &       &  & \phterm{i-2}{i-1}, &  & \phterm{i-2}{i}, &  & \phterm{i-2}{i+1}, &  & \dots          \\
	               &       &  &                    &  & \phterm{i-1}{i}, &  & \phterm{i-1}{i+1}, &  & \dots          \\
	               &       &  &                    &  &                    &  & \phterm{i}{i+1}, &  & \dots          \\
	               &       &  &                    &  &                    &  &                  &  & \dots
					\},
\end{align*}
\noindent where the $i$-th row shows prediction data from $t_i$ to all the following epochs $t_j$ represented by the $j$-th columns.

We choose these learning variables and target variables based on the rationale that the true state and the true models are usually not available, so any variables that explicitly depend on the true orbit information shall be not used, such as the true error $\bm{e}$. 
However, the information of the true error $\bm{e}$ can be implicitly contained in the relative error $\bm{\varepsilon}$ to some degrees. 
In this way, the proposed ML approach actually also connects the predicted error with the true prediction error at a future epoch, with the assistance of other learning variables.

\section{Numerical Results of Improving Orbit Prediction}
\label{sec:numerical results}

In this section, the framework established in the previous section is applied to a simulated RSO. 
The ML approach to improve orbit prediction is demonstrated by training an SVM model first and then using it to improve the orbit prediction accuracy. 
We will also discuss the effect of the size of the training data at the last subsection. 

\subsection{Building the SVM model for the Simulated RSO}

Parameters for the simulated RSO are summarized in \cref{tab:RSO_parameters}. The initial orbital parameters are given in the ECI frame, which is directly converted from a TLE of the International Space Station (ISS). 
However, we note here that we study a generic RSO with area-to-mass ratio and reflection coefficient different from the ISS. 

%The parameters for the simulated RSO in this section is summarized in . 
%The initial orbital parameter is given in the ECI frame. 
%It is directly converted from a TLE data of the International Space Station (ISS), which indicates that the simulated RSO is assumed to be a space debris with potential hazard to the safety of the ISS. 

% \usepackage{booktabs}
\begin{table}[!htbp]
	\centering
	\caption{Parameters of the simulated RSO.}
	\label{tab:RSO_parameters}
	\footnotesize
	\begin{tabular}{ll}
		\toprule
		Parameter                          & Value               \\ \midrule
		UTC epoch                          & 2016/09/28 12:09:45 \\
		Semi-major axis $a$ [km]           & 6783.34             \\
		Eccentricity $e$                   & 0.006793            \\
		Inclination $i$ [deg]              & 51.6393             \\
		Argument of perigee $\omega$ [deg] & 14.5438             \\
		RAAN $\Omega$ [deg]                & 262.6471            \\
		Mean anomaly [deg]                 & 345.5909            \\
		Area-to-mass ratio                 & 0.05                \\
		Drag coefficient $C_d$             & 2.2                 \\
		Reflection coefficient $C_r$       & 1.25                \\ \bottomrule
	\end{tabular}
\end{table}

The RSO is propagated for 30 days using the truth dynamic model. 
The numerical integrator is chosen to be the Dormand-Prince 8(5,3) method for all the propagation in this paper. 
The absolute tolerance on each position component in the ECI frame is set as 0.1 m, and the propagation step size is limited between 0.001 and 90 seconds. %, and the velocity accuracy is set to ${(\mu_{\rm Earth} \delta r)}/{(v r^2)}$ accordingly, which is set up using the Orekit. 
%~\footnote{At its core, DOP is a Dormand \& Prince 8th order method that uses a 5th order error estimator and bootstraps on a 3rd order estimator to obtain a dense output of order 7.~\citep{DP853}} 

\begin{figure}[!htb]
	\centering
	\includegraphics[width=0.45\textwidth]{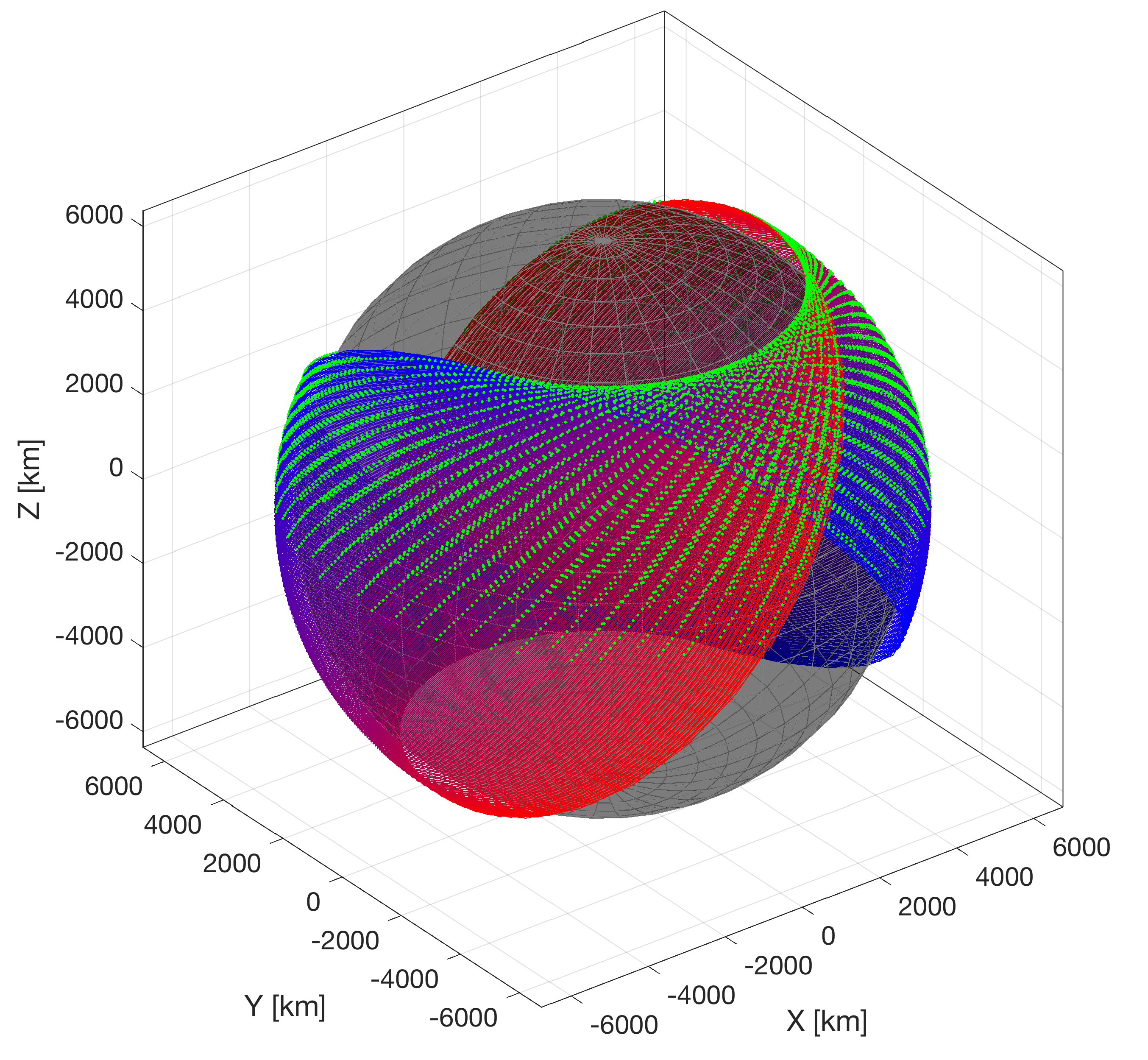}
	\caption{True orbit (colorful arcs) and observations (green dots) of the RSO during the simulation in the ECI frame. The true orbit is colored by red at the beginning, and gradually transits to the blue color at the end of the propagation.}
	\label{fig:orbit_3d}
\end{figure}

\begin{figure*}[!htb]
	\centering
	\includegraphics[width=0.9\linewidth]{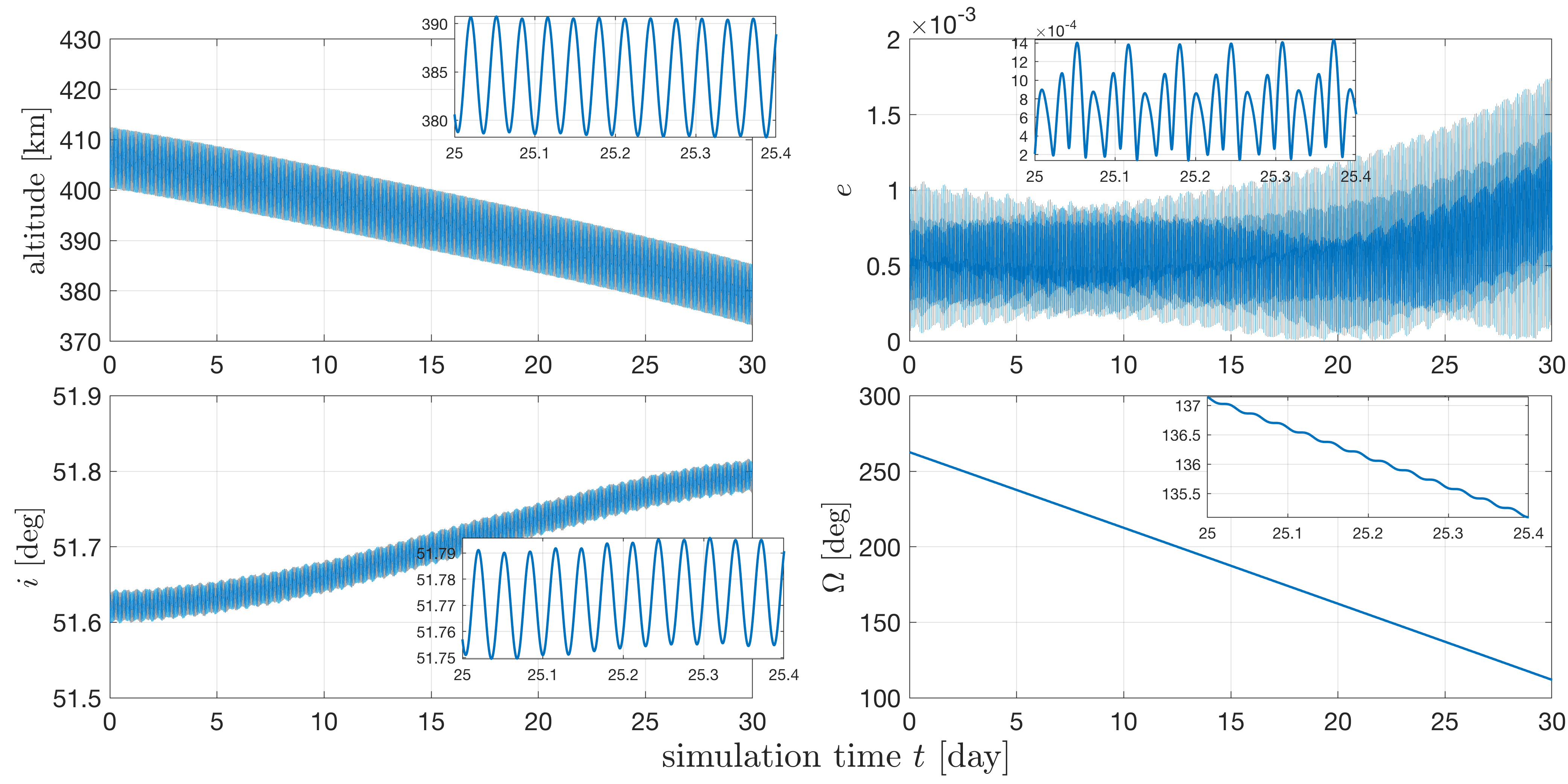}
	\caption{Instantaneous classical orbital elements of the simulated RSO.}
	\label{fig:orbit_coe_curve}
\end{figure*}

\Cref{fig:orbit_3d} demonstrates the orbit of the simulated RSO for 30 days. 
The orbit is colored by red at the beginning, and is seen to gradually change to blue at the end. This is expected as that the orbit processes within the simulated time interval, due to the perturbation forces. 
In \cref{fig:orbit_3d}, the green dots along the orbit show the locations where measurements are generated by the ground stations. 
The revolution of the instantaneous COE is shown in \cref{fig:orbit_coe_curve}, alongside with zoom-in plots in $t\in[25,25.4]$ days to clearly show their variations. 
Apart from the clear precession motion, the orbit also shows short-term and secular changes on the semi-major axis, eccentricity and inclination. 
%Because the orbit of the RSO is nearly circular, its perigee argument $\omega$ is sensitive to the, so as the true anomaly $f$. They are also continuously varying but not demonstrated in \cref{fig:orbit_coe_curve}.  Other two components of the COE is omitted because 
%The simulation results show that the motion of the RSO greatly depends on the 

After obtaining the dataset following the description in the previous section, we perform data cleaning before starting to train the SVM model. 
The guidelines for the data cleaning usually depend on the specific dataset, the target problem, and also the experience of the modeler. 
Here, two types of data cleanings are performed. 
First, we remove the cases where the estimations of the drag coefficient $\hat{C}_d$ lie outside of the interval $[1.2,3.2]$. Because the atmosphere drag is sensitive to $\hat{C}_d$, a largely biased $\hat{C}_d$ can cause significant errors. 
In practice, $\hat{C}_d$ can be compared with the historical data, and those with large biases can be detected. %, and such estimations should be treated as mistakes rather a regular estimation error. 
Second, we use the Tukey's test~\citep{tukey_exploratory_1977} to remove outliers in the dataset, which means data points with too large errors are eliminated. 
Specifically, denoting the quartiles of the along-track error $e_y$ as $q_1$, $q_2$, and $q_3$, the errors out of the interval $[ q_1 - 3 (q_3-q_1) ,\allowbreak  q_1 + 3 (q_3-q_1) ]$ are discarded as outliers.  
%We remark that it is possible that they might be caused by very biased predictions rather than truly outliers. 
%However, we assume that the ML approach will only deal with regular orbit prediction errors in this study, which should be neither too large nor too small. 
%For other situations, there are more appropriate and conventional methods. 

In this paper, the MATLAB (R2017b) implementation of SVM regression function \texttt{fitrsvm} is used~(MathWorks, \citeyear{the_mathworks_statistics_2017}). 
The kernel function is set to the Gaussian kernel to deal with the nonlinearity of the prediction error. 
Some default settings of the MATLAB implementation are used: 
the scaling parameter for the Gaussian kernel is selected by a heuristic procedure with a subsampling process; 
the $\varepsilon$-margin is selected as ${\rm iqr}(\bm{y})/13.49$ where ${\rm iqr}(\bm{y})$ indicates the interquartile; 
the box constraint parameter $C$ is selected as ${\rm iqr}(Y)/1.349$; 
the optimization solver for the SVM training is the Sequential Minimal Optimization (SMO) method. 
The learning variables will be standardized using the mean and standard deviation of each variable. 
The Karush-Kuhn-Tucker (KKT) violation is checked at every iteration to determine the convergence when solving the optimization problem, where the convergence tolerance is $10^{-3}$. 
Based on the collected dataset, six SVM models in total can be trained to capture the underlining orbit prediction errors for the six components of the true error ${}^{\mathsf{RSW}}\delta\bm{x}_{\bm{e}} = [e_x,\allowbreak e_y,\allowbreak e_z,\allowbreak e_{vx},\allowbreak e_{vy},\allowbreak e_{vz}]^{\rm T}$. 

Additionally, in order to reproduce the training results, the seed of the random stream is set to 42 before the training. 
We remark that our experiments reveal that the results in the following discussions do not depend on the particular random seed. 
A different seed will slightly vary the specific metric but will not change the general pattern.

\subsection{Validation of the Trained SVM Model}

First, we consider a simpler case, where only the first ground station in \cref{tab:stations} is considered. 
The measurements and estimation errors of a single station are more likely to be coherent, thus the underlining prediction errors are more likely to follow a deterministic pattern, which can be learned by the SVM. 
This case reflects the situation where a station only has access to its own historical data. 

\begin{figure}[!htbp]
	\centering
	\subfigure[Position components (logarithmic plot of absolute value)\label{subfig:position error}]{\includegraphics[width=0.48\textwidth]{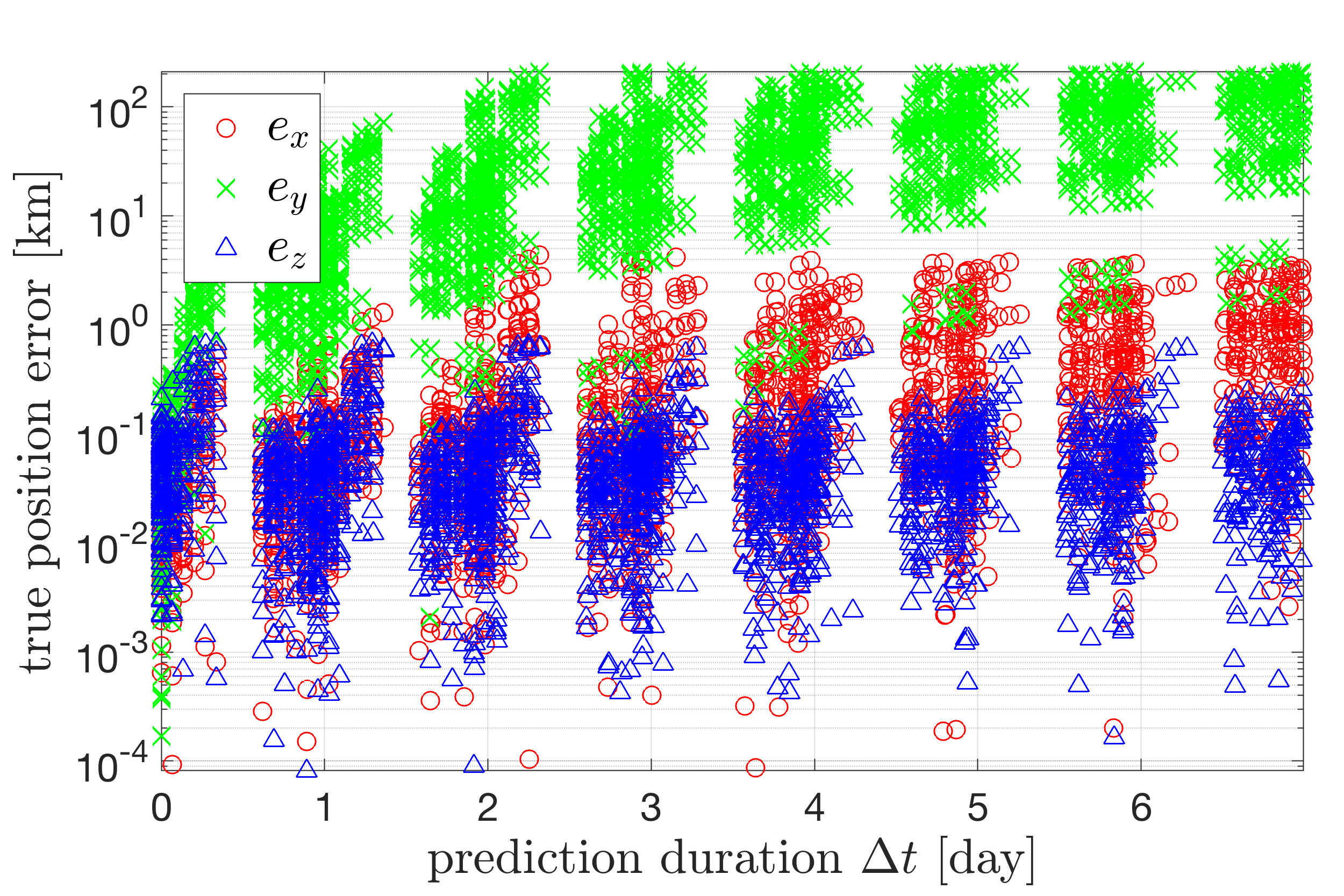}}
	\subfigure[Velocity components (real value)\label{subfig:velocity error}]{\includegraphics[width=0.48\textwidth]{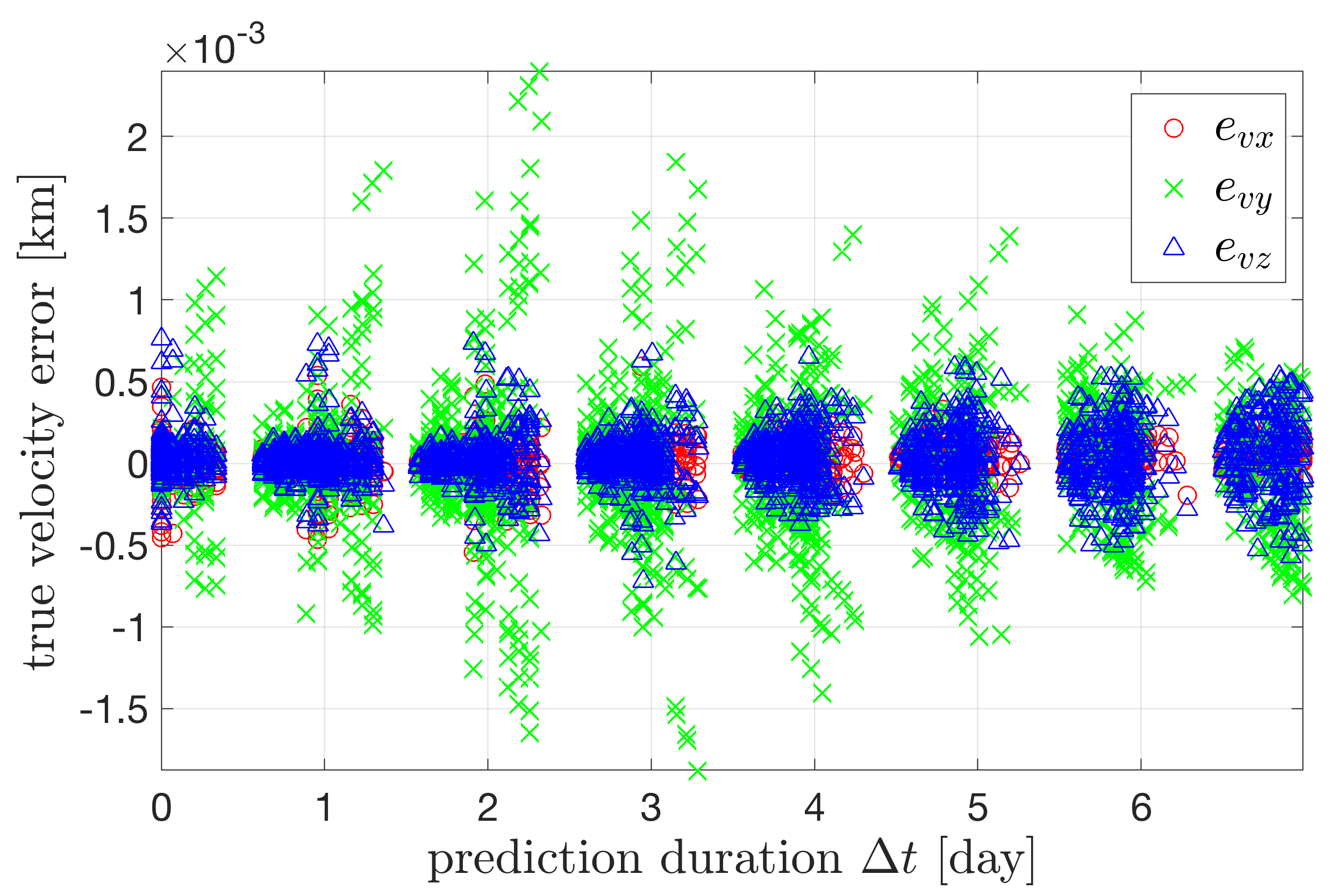}}
	\caption{Distribution of the true prediction error $\bm{e}$ with respect to the prediction duration $\Delta t$.}
	\label{fig:position and velocity error (1 station)}
\end{figure}

\Cref{fig:position and velocity error (1 station)} shows the true prediction error $\bm{e}$ of the whole dataset, where the RSO has been propagated for 30 days and the maximum prediction duration $\Delta t_{\rm max}$ for each estimated state is set is 7 days. 
The horizontal axis represents the prediction duration $\Delta t$. 
The vertical axis represents the position components of $\bm{e}$ in \cref{subfig:position error} and the velocity components in \cref{subfig:velocity error}. 
In \cref{subfig:position error}, the position errors are shown in the logarithm coordinate, since the along-track position error $e_y$ is significantly larger than the other two components. 
In \cref{subfig:velocity error}, although $e_{vy}$ is slightly more spread out than the other two components, all the errors are small and there is no dominant component. 
The most important task is to reduce the orbit prediction error of the along-track position error $e_y$. % and at the same time reduce or at least not worse the other components of the orbit prediction error. 
Therefore, in the following study, we will only use $e_y$ as the study object and demonstrate the performance of reducing $e_y$ through the ML approach.

80\% of the dataset is randomly chosen to train the SVM model while the remaining 20\% is used to evaluate the performance of the learned model.
For evaluation, the learning variable $\bm\Lambda$ of the testing data is the input to the SVM model, and the output is the ML-predicted error, denoted as $\hat{\bm{e}}_{\rm ML}$. 
The ML-modified prediction $\hat{\bm{X}}_{\rm ML}$ is achieved by subtracting $\hat{\bm{e}}_{\rm ML}$ from the prediction $\hat{\bm{X}}$. 
So the residual orbit prediction error $\bm{e}_{\rm res}$ after the ML-modification is 
\begin{equation} \label{eq:residual error}
	\bm{e}_{\rm res} = \bm{e} - \hat{\bm{e}}_{\rm ML} = \hat{\bm{X}}_{\text{ML-modified}} - \bm{X},
\end{equation}
which shall be zero if the SVM model completely captures the underlying errors. 

We use the concept of the volume-weighted symmetric mean absolute percentage error (volume-weighted MAPE) from statistics as the metric to evaluate the prediction accuracy between the reference value and the output value of a prediction model. 
The performance metric $P_{{\rm ML}}$ used to quantify the trained SVM model is defined as the ratio between the summation of the absolute errors between $e$ and $e_{\rm ML}$ on each data point in the testing data, and the summation of the original absolute true error of all data in the testing data, which can be mathematically expressed as,
\begin{equation} \label{eq:SMAPE}
	P_{{\rm ML}} = 100\% \cdot \frac{\sum_{i=1}^n |e-\hat{e}_{\rm ML}|}{\sum_{i=1}^n |e|},
\end{equation}
where $n$ is the size of the testing data. 
The metric reaches its lower boundary zero when the ML-predicted error is identical to the true error, but has no upper boundary.
%The absolute error between the true error $\bm{e}_{1}$ and the ML-predicted error $\hat{\bm{e}}_{\rm ML}$ on each data point are summed up, and then divided by the total absolute true error of the testing data. 
%The metric has a lower boundary of 0 when the ML-predicted error is identical to the true error, but has no upper boundary. 
Notice that the numerator in \cref{eq:SMAPE} is actually the residual error after ML-modification, as defined in \cref{eq:residual error}. 
Therefore, we have
\begin{equation}
	P_{{\rm ML}} = 100\% \cdot \frac{\sum_{i=1}^n |e_{\rm res}|}{\sum_{i=1}^n |e|},
\end{equation}
which indicates that the metric actually measures the percentage of the residual error $\bm{e}_{\rm res}$ with respect to the true error of the testing data. 
The lower this percentage $P_{\rm ML}$ is, the more errors are corrected, thus the better the performance of the trained SVM model is. 

Notice that this chosen metric is consistent with the objective of the SVM model, which is to include all data point with a surface as smooth as possible through minimizing the total error within the training data. In the following analysis, this metric will be used to quantify the performance of the trained SVM model. 
%Moreover, the volume-weighted MAPE is consistent with the object of the SVM model, because the algorithm of the SVM is trying to include all data point with a surface as smooth as possible, which is actually to minimize the total error within the training data. 
%%
%In the following analysis, this metric will be used to quantify the performance of the trained SVM model. 
%However, attentions should be paid that this metric depends on particular testing data and can be notably effected by some extreme outliers, even though we have tried to only study regular dataset. 
%And we suggest that in most cases, the tendency rather than the particular value of this metric should be concerned. 

\begin{figure}[!htb]
	\centering
	\subfigure[Absolute value.\label{subfig:training result 80 20 1 station absolute}]{\includegraphics[width=0.9\linewidth]{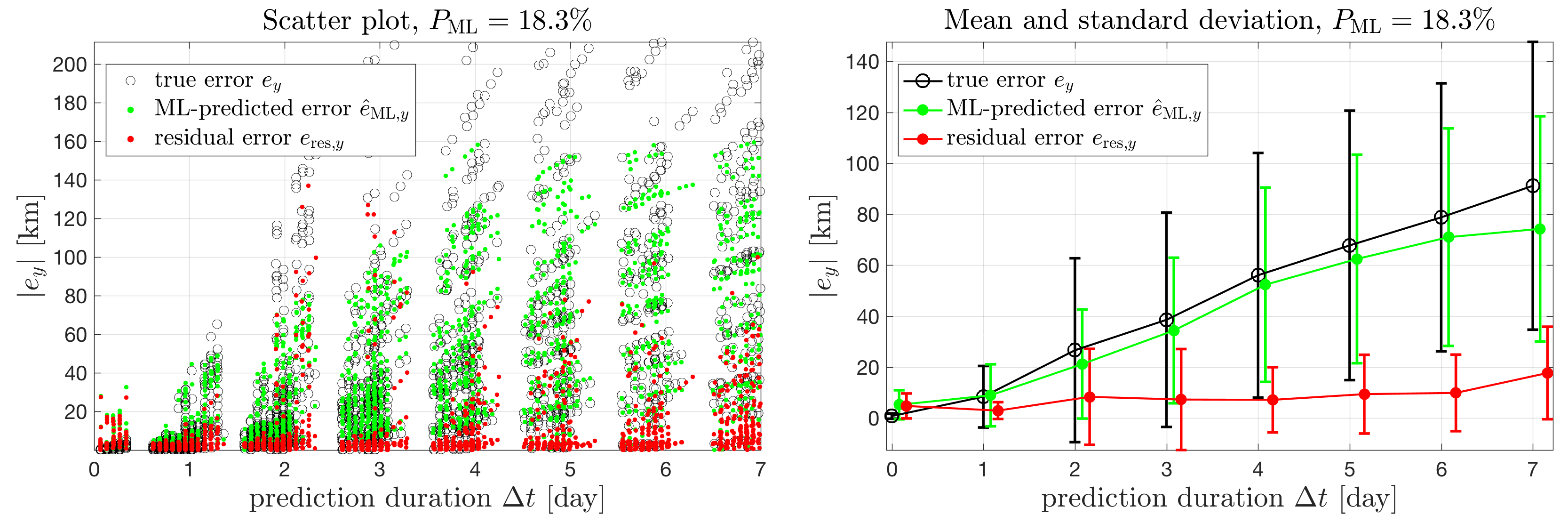}} \\
	\subfigure[Real value.\label{subfig:training result 80 20 1 station real}]{\includegraphics[width=0.9\linewidth]{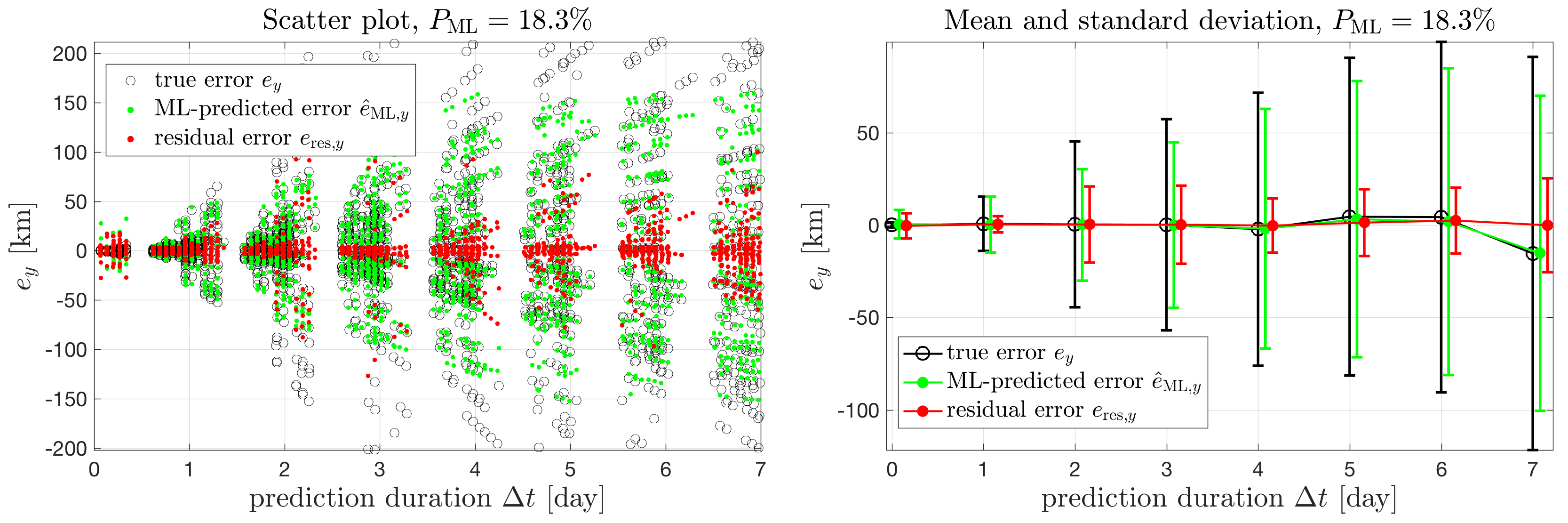}}
	\caption{Results of the trained SVM model on training data with 1 station.}
	\label{fig:result on training data 1 station}
\end{figure}

\Cref{fig:result on training data 1 station} demonstrates the results of the SVM model on the training data. 
The horizontal axis represents the prediction duration $\Delta t$. 
The vertical axis shows true, ML-predicted and residual along-track errors $e_y$ respectively, as annotated by the legend in the figure. 
In both \cref{subfig:training result 80 20 1 station absolute,subfig:training result 80 20 1 station real}, the left figures directly show the scatter plot, where the black circle represents the true error $e_{y}$, the green dot represents the ML-predicted error $\hat{e}_{{\rm ML},y}$, and the red dot represents the residual error $e_{\rm res}$ after ML-modifications. 
The right figures show the errorbar plot of the left scatter plot, where the center point represents the mean value of each clustered group of the error, and the length from the middle of the bar to the top (or the bottom) represents the standard deviation of the corresponding clustered group. 
For clarity, the three curves are slightly displaced along the horizontal axis to avoid overlapping. 
(Similar figures will appear in the remaining part of the paper, and we will only explain the difference there.)

In \cref{subfig:training result 80 20 1 station absolute}, the absolute values of the errors are demonstrated. 
The ML-predicted error $\hat{e}_{{\rm ML},y}$ is scattering within the same area of the true error $e_{y}$, and the errorbar curve is very close to each other. 
The performance metric $P_{\rm ML}$ is 18.3\%, indicating that the trained SVM model has captured most features of the underlying error model. 
The residual error is also very small as revealed by both the scatter plot and the errorbar plot. 
Therefore, we can conclude that the trained SVM model captures the orbit prediction error in the training dataset very well. 
In \cref{subfig:training result 80 20 1 station real}, learning results are demonstrated with the real value of the errors. 
As shown in both plots, the error is distributed around zero almost evenly. 
In this case, the orbit prediction error cannot be compensated for if we only perform a modification based on fitting the mean error with respect to the prediction duration $\Delta t$, which was reported in the reference~\citep{rivera_improving_2016}. 
In contrast, the trained SVM model successfully compensates for the errors and reduces the standard deviation. 
%Meanwhile, the mean value is still kept around zeros or even improved at $\Delta t=7$ days.

%In the above analysis, since the training data is directly taking as the testing input, the remaining error should be the randomness introduced during the measurement and estimation procedures, and also some residual errors beyond the capability of the SVM model or not covered in the learning variables discussed in the previous section. 
%%However, it could arise from many different resources such as un-included variables in the machine learning dataset. 
%Since the percentage of the residual error $P_{\rm ML}$ is small, we can conclude that the SVM model can captures the hidden pattern of the training data. 

\begin{figure}[!htb]
	\centering
	\subfigure[Absolute value. \label{subfig:scattering plot 2}]{\includegraphics[width=0.9\linewidth]{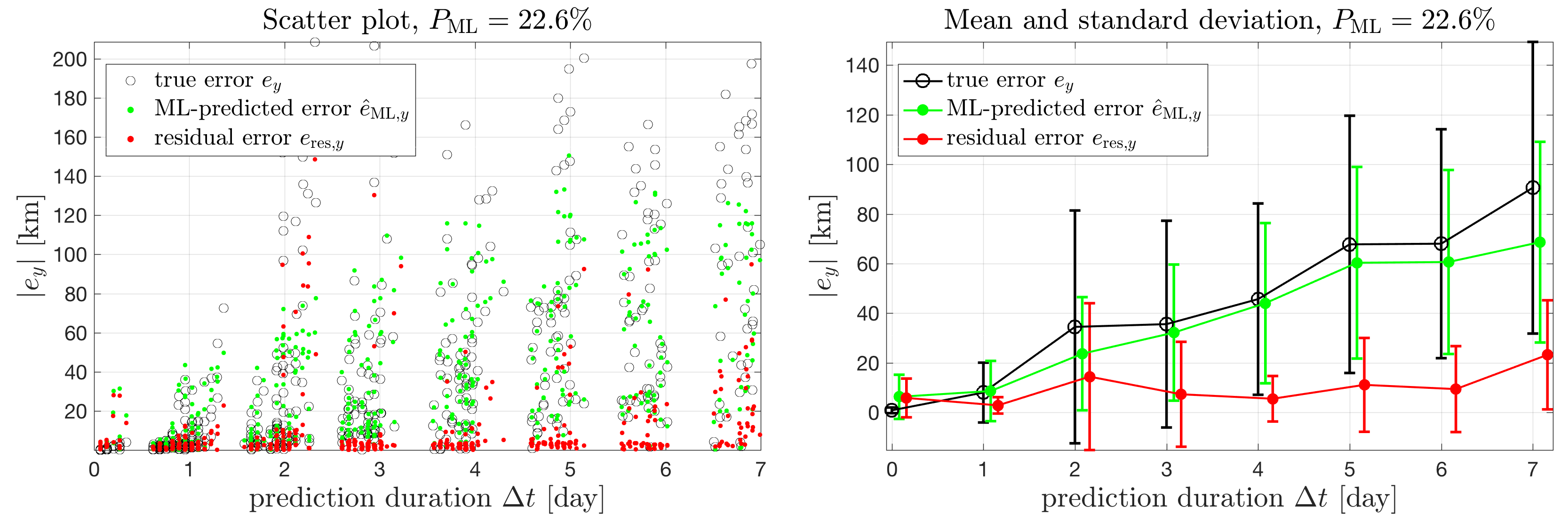}} \\
	\subfigure[Real value. \label{subfig:errorbar plot 2}]{\includegraphics[width=0.9\linewidth]{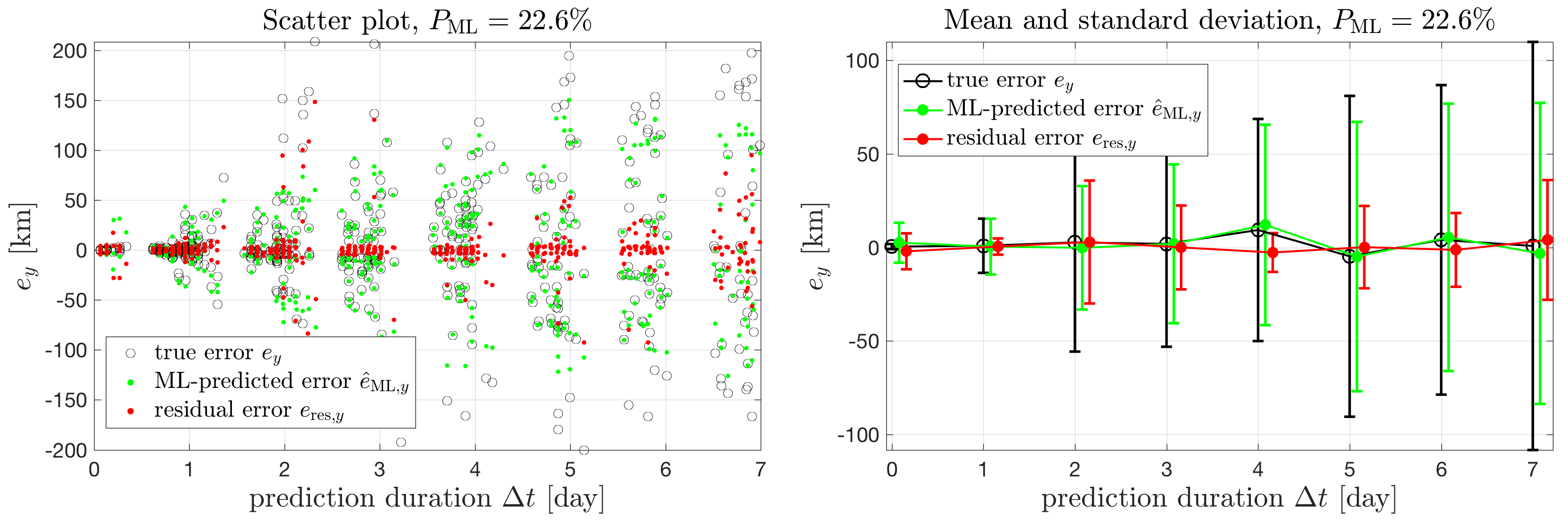}}
	\caption{Results of the trained SVM model on testing data with 1 station.}
	\label{fig:result on testing data 1 station}
\end{figure}

A common concern in the machine learning applications is the performance of the trained model on a different set of data that is not included in the training data. 
We test this generalization capability here, where the remaining testing data is used as the input to the trained SVM model, and the results are shown in \cref{fig:result on testing data 1 station}. 
The metric $P_{{\rm ML}}$ is 22.6\% now, which is larger than that of the training data. 
This is reasonable because the testing data can contain information beyond the training data. 
The errorbar plots in both subfigures show that the mean value and the standard deviation have been greatly reduced, even though the testing data is unknown to the trained SVM model. 

\begin{figure}[!htb]
	\centering
	\subfigure[Absolute value]{\includegraphics[width=0.9\linewidth]{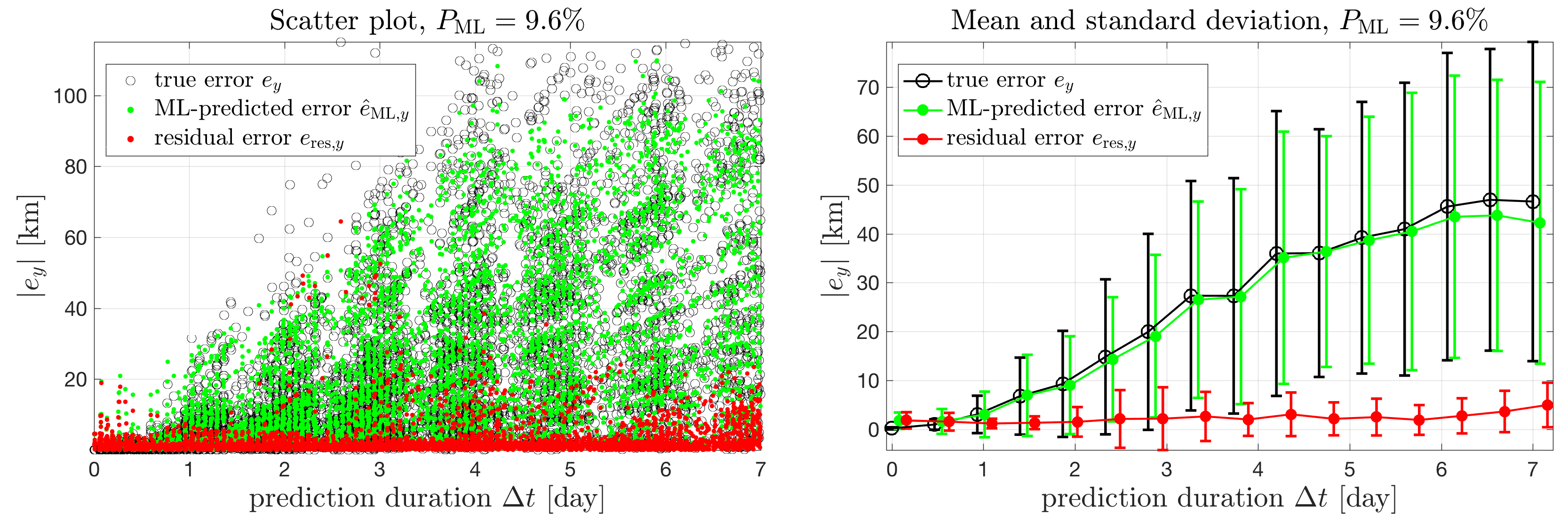} }
	\subfigure[Real value]{\includegraphics[width=0.9\linewidth]{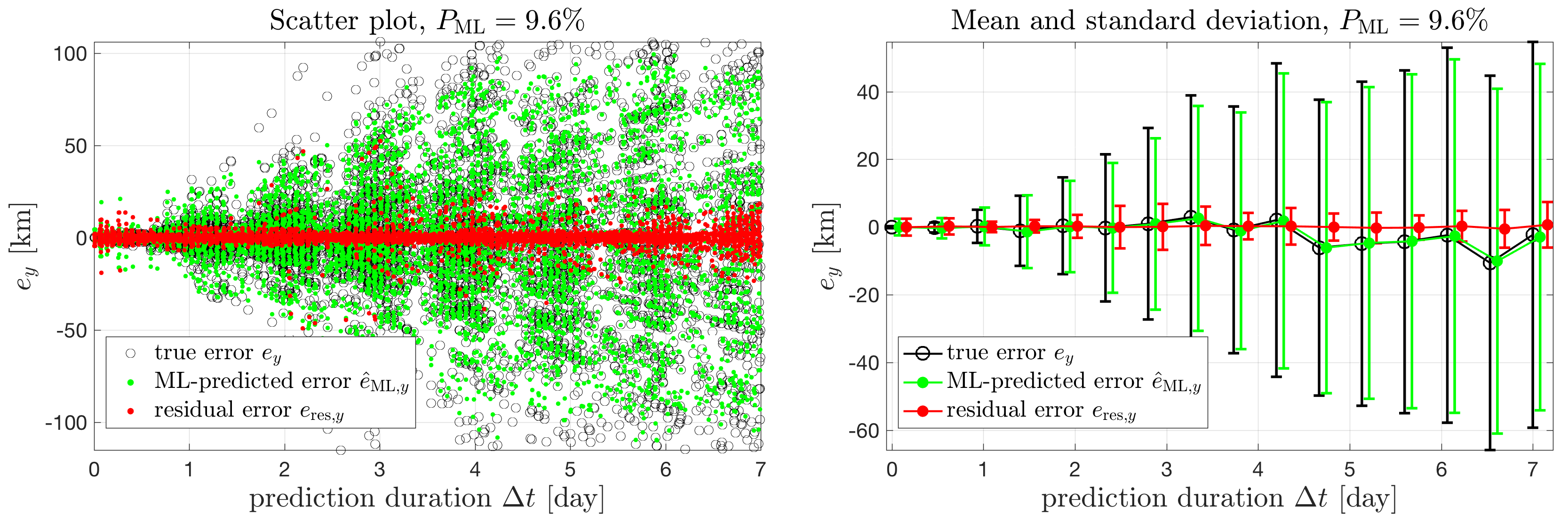} }
	\caption{Results of the trained SVM model on testing data with 3 stations.}
	\label{fig:result on training and testing data 3 stations}
\end{figure}

In the above discussions, only one station is considered. 
A clear drawback of this situation is that a RSO is possible to be below the horizon of the station for a long time, as revealed by the gaps in \cref{fig:result on testing data 1 station,fig:result on training data 1 station}. 
Next, all three ground stations in \cref{tab:stations} are included in the simulations. 
The obtained dataset is also randomly (with fixed seed to allow reproduction) divided into training (80\%) and testing (20\%) data. 
The result of the new SVM model on the testing data is shown in \cref{fig:result on training and testing data 3 stations}. 
We observe two clear differences between the new results and the results using only one station: 
\begin{itemize}
	\item The data in the scatter plot is denser along the horizontal axis, due to the fact that more tracks are available when three stations are used. The results have been clustered into more groups in the error bar plots to show more details in the figure. 
	\item The maximum of true prediction errors (black circles) are smaller, due to the fact that the estimations become more accurate when more measurements are available. 
\end{itemize}
The performance has been greatly improved when the three stations are used. 
The metric $P_{{\rm ML}}$ is 9.6\% now, improved by more than a half compared with that of the single station situation, where $P_{\rm ML}$ is 22.6\%. 
The mean errors have been reduced to almost zero, and the standard deviations have also been greatly reduced.

As revealed in the above analysis, with more available ground stations, the trained SVM model can capture the underlying orbit prediction error better, and the generalization performance is also significantly improved. 
In the following studies, we will only present results when three ground stations are used.

\subsection{Different Partition of Dataset}

An often accepted concept for the machine learning method is that the more training data used, the better performance the machine learning model has. 
In this subsection, we study this effect by investigating different partition of the dataset on the performance of the trained SVM model. \Cref{fig:different partition} shows the result of the trained SVM model, using 40\% or 90\% of the dataset as training data respectively. 
(Note in the follow analysis, for clarity, only the absolute value of the result are demonstrated.)
The performance is clearly improved as shown by the thiner and steadier errorbar curves of the residual error $e_{\rm res}$. 
Also, the metric $P_{{\rm ML}}$ drops from 14.8\% to 9.7\%. 

\begin{figure}[!htb]
	\centering
\newcommand{\tempOne}[2]{\subfigure[#1]{\includegraphics[trim={22cm 0 0 0}, clip=true, width=0.45\linewidth]{#2} }}
	\tempOne{training data is 40\% of dataset}{"Figures/sec43/e_y-day0-30-ratio40,60-testing-absolute"} 
	\tempOne{training data is 90\% of dataset}{"Figures/sec43/e_y-day0-30-ratio90,10-testing-absolute"} 
	\caption{Results of the SVM model trained by different size of training data.}
	\label{fig:different partition}
\end{figure}

\begin{figure}[!htb]
	\centering
	\includegraphics[width=0.7\linewidth]{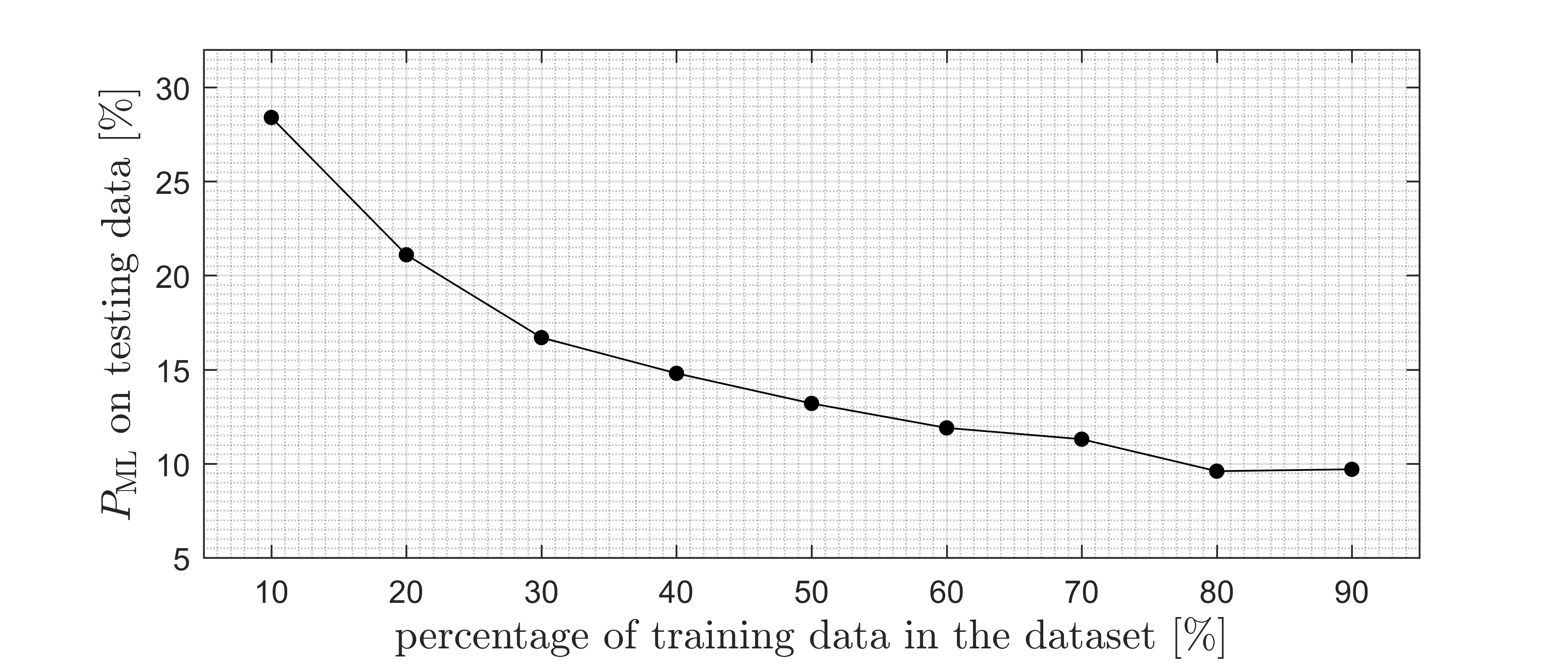}
	\caption{Performance metric $P_{\rm ML}$ of the SVM model trained by different size of training data.}
	\label{fig:curve_different_partition}
\end{figure}

The variation of the metric with respect to the size of the training data is further investigated by varying the percentage of the training data from 10\% to 90\%. 
The results are shown in \cref{fig:curve_different_partition}. 
Interestingly, we notice that the metric $P_{\rm ML}$ drops quickly at the beginning, but tends to a constant once the size of the training data exceeds 80\%. 
This implies that the performance of the SVM model will not be further improved once sufficient amount of training data has been used. 
Additionally, the results validate that our choice of 80\% dataset in the previous subsections is appropriate to achieve quality performance.

\section{Generalize the SVM Model to Future Epochs}
\label{sec:generalize to future}

In the previous section, the generalization capability of the trained SVM model is evaluated similarly as the classical machine learning studies where the ML model performance is tested using testing data different from the training data, without distinguishing in which way the testing data is different from the training data. 
In improving the orbit prediction of RSOs, we are more interested in testing the SVM model for future epochs. 
Therefore, in this section, we introduce a constraint on the testing data, requiring them to be predictions at future epochs with respect to the training data. 

\subsection{Results on Simulated RSO}

The simulation of the RSO is extended to 50 days, in order to provide testing data beyond the training data within the first 0--30 days. 
The data in the first 30 days is used to train the SVM model, the data in the 30--40 days (future 10 days) is collected as the testing data of the trained SVM model, other data are used to analyze the dependency of the performance on the amount of the testing data. 
Again, as the along-track error $e_y$ is the largest component in the RSW frame, we will only demonstrate the performance of the trained SVM model on $e_y$. 

\begin{figure}[!htbp]
	\centering
	\subfigure[Absolute value] {\includegraphics[width=0.9\linewidth]{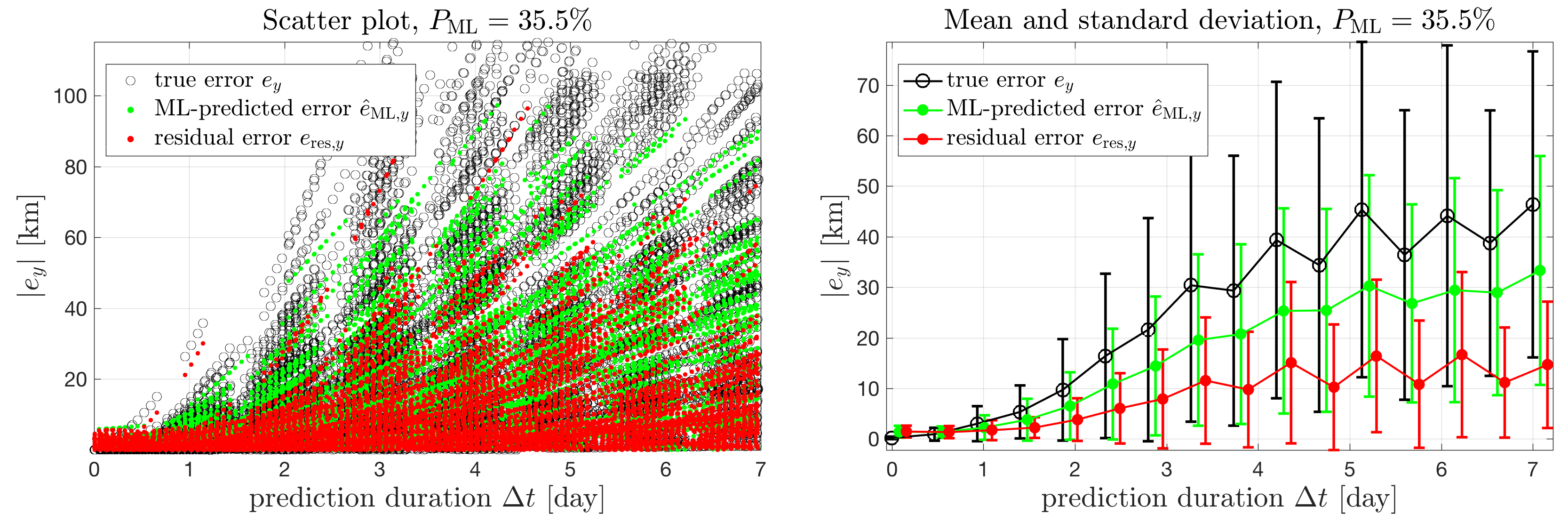}} \\
	\subfigure[Real value] {\includegraphics[width=0.9\linewidth]{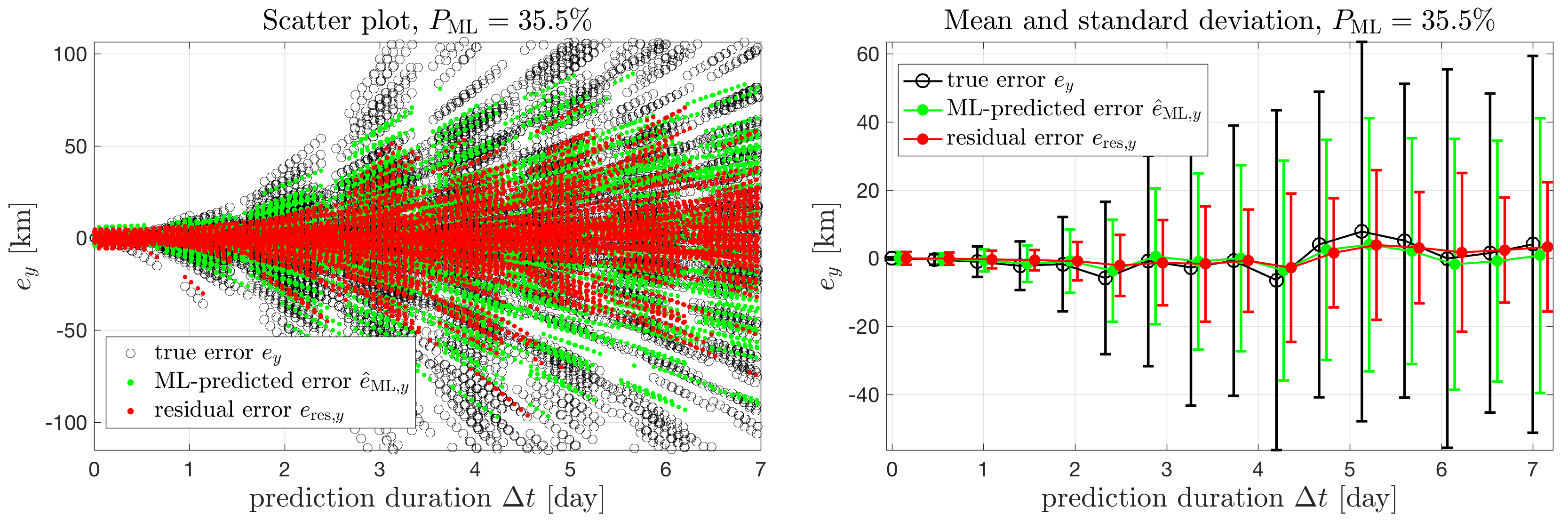}}
	\caption{Results of using the SVM model trained by data in 0--30 days, to modify the error in 30--40 days (future 10 days).}
	\label{fig:along track error}
\end{figure}

\Cref{fig:along track error} shows the results using training data in the first 0--30 days to modify predictions in the following 10 days (30--40 days). 
In the plot of the absolute value, the mean error is reduced from around 50 km to less than 20 km at $\Delta t=7$ days. 
The standard deviation is reduced by more than a half in most cases. 
The plot of the real value implies that the error is distributed almost evenly along both positive and negative directions, and the mean value is reduced to almost zero with less fluctuations. 
The metric $P_{{\rm ML}}$, the percentage of the residual error, is reduced to 35.5\% of the true error. 
Compared with the results in the last section, in \cref{fig:result on training and testing data 3 stations}, the performance is worse as the metric becomes larger, which is expected since generalizing the trained SVM model to the future time should be harder than within the same time interval. 
Fundamentally, the orbit of the studied RSO has changed, as illustrated in \cref{fig:orbit_3d,fig:orbit_coe_curve}.

We note that the maximum prediction duration $\Delta t_{\rm max}$ in \cref{fig:along track error} is 7 days, and the predictions in the testing data are made at epochs in the following 10 days. 
For example, a prediction made at $t=40$ days can base on the estimation given at $t=33$ days at the earliest (with $\Delta t = \Delta t_{\rm max} = 7$ days), and in this situation both the prediction and its based estimation are not included in the training data. 
So there is no contradiction between $\Delta t_{\rm max}$ and time length of the testing data. 

\begin{figure}[!htbp]
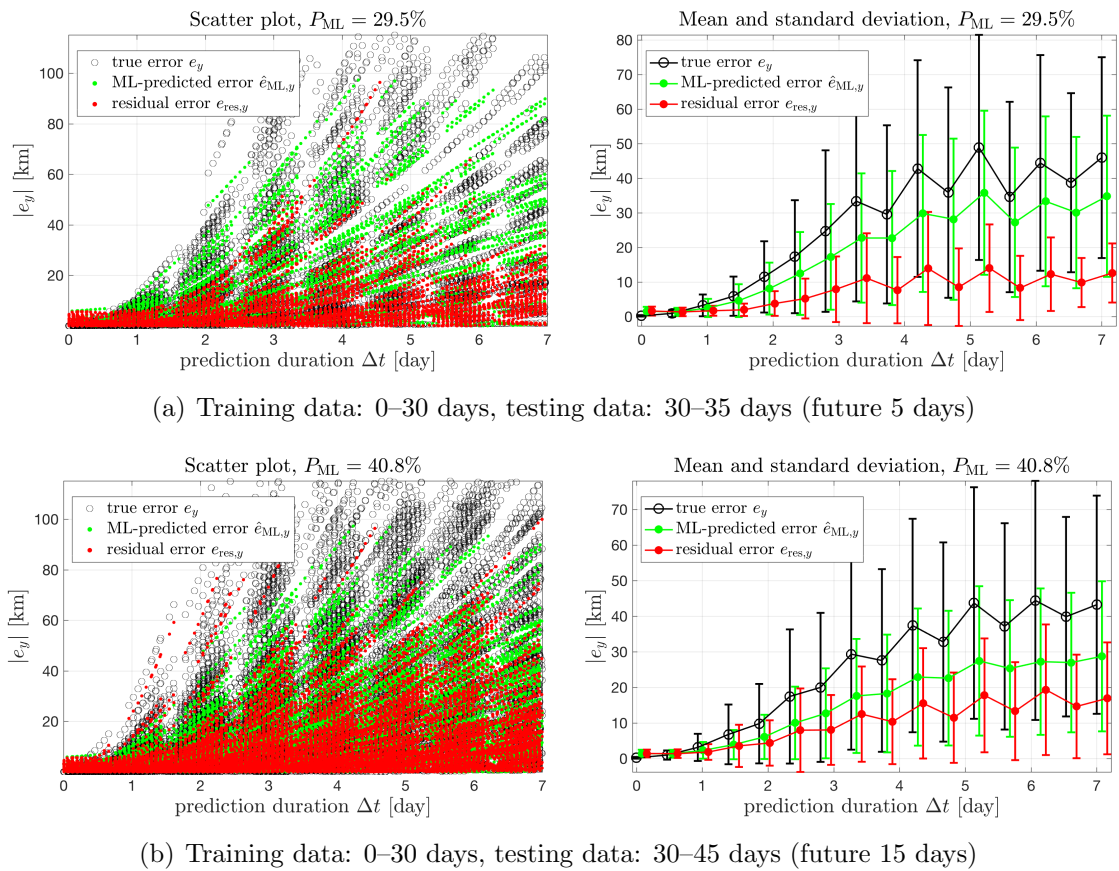

	\centering
	\newcommand{\tempOne}[2]{\subfigure[#1]{\includegraphics[trim={0cm 0 0 0}, clip=true, width=0.9\linewidth]{#2}}}
	\tempOne{Training data: 0--30 days, testing data: 30--35 days (future 5 days)\label{subfig:future 5 days}}{"Figures/sec51/e_y-day0-30-30-35-testing-absolute"}
	\tempOne{Training data: 0--30 days, testing data: 30--45 days (future 15 days)\label{subfig:future 15 days}}{"Figures/sec51/e_y-day0-30-30-45-testing-absolute"}
	\caption{Results of the trained SVM model on different length of future epochs.}
	\label{fig:different length of testing data}
\end{figure}

We show other two cases, with the testing data in future 5 days and 15 days respectively, in \cref{fig:different length of testing data}. The top one is for the testing data in 30--35 days, and the bottom one is for 30--45 days. 
Comparing \cref{subfig:future 5 days,subfig:future 15 days}, we observe that the percentage $P_{\rm ML}$ of residual error after ML-modification increases as further-into-future testing data is used.

%\subsection{Individual Performance at Future Epochs}
So far, we have demonstrated that the average performance evaluated by $P_{\rm ML}$ can be improved by the proposed ML approach. 
In the practice, it is also possible that the individual performance of the trained SVM model on one data point at a future epoch is concerned. 
The ideal case is that all the residual errors after the ML-modification are smaller than the true prediction errors. 
To evaluate the individual performance of the trained SVM model, we introduce an individual metric $p_{\rm ML}$, defined with a given $\bm{\Lambda}_i$ as
\begin{equation}
	{\left. p_{{\rm ML}} \right|_{\bm{\Lambda}_i}} 	= 	{\left. 100\% \cdot \frac{|e-\hat{e}_{\rm ML}|}{|e|} \right|_{\bm{\Lambda}_i}} 	=	{\left. 100\% \cdot \frac{|e_{\rm res}|}{|e|} \right|_{\bm{\Lambda}_i}} . \\
\end{equation}
The lower bound of $p_{\rm ML}$ is 0 when the orbit prediction error is perfectly compensated, while the upper bound of $p_{\rm ML}$ is positive infinity. 

\begin{figure}[!htbp]
	\centering
	\subfigure[Data point with reduced errors.\label{subfig:reduced}]{\includegraphics[width=0.49\linewidth]{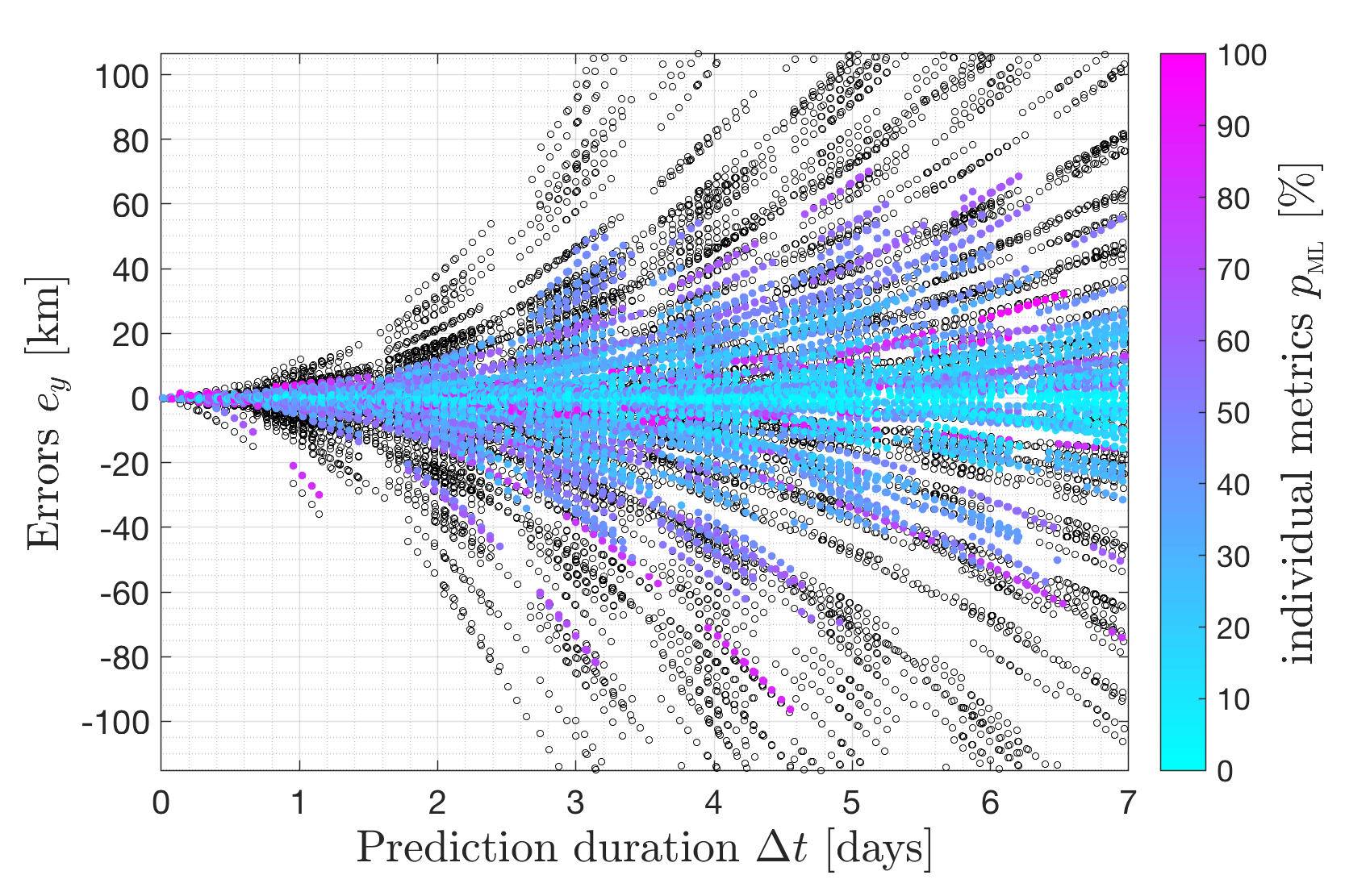}}
	\subfigure[Data point with increased errors.\label{subfig:increased}]{\includegraphics[width=0.49\linewidth]{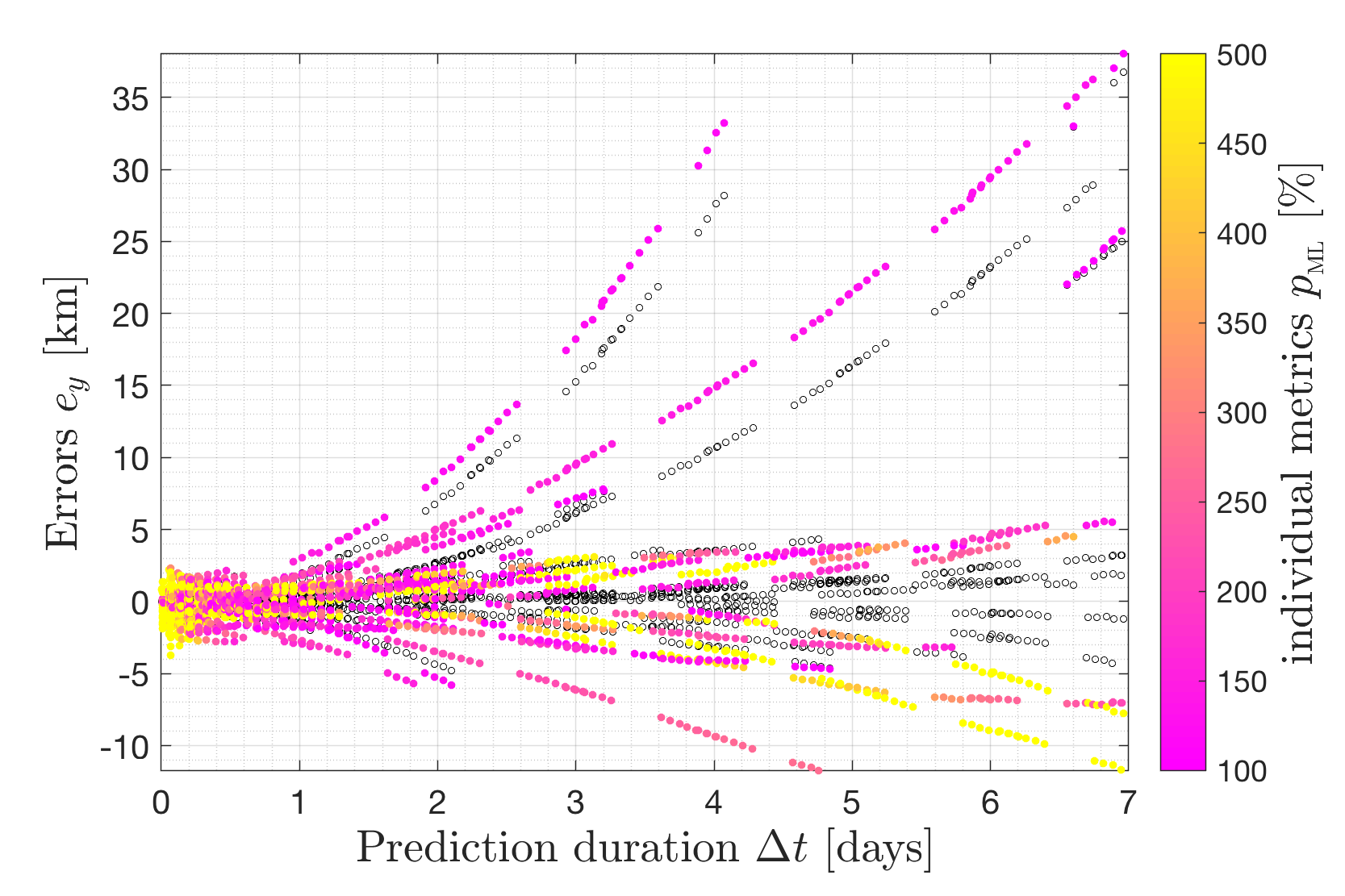}}
	\caption{Individual performance of the trained SVM model on future epochs, demonstrated by true error (black circles) and residual errors after ML-modification (colored by $p_{\rm ML}$).}
	\label{fig:LEO1 colored by pML}
\end{figure}

The performance of the trained SVM model on the simulated RSO, using the same data as in \cref{fig:along track error}, is demonstrated in \cref{fig:LEO1 colored by pML}, with only the true error (black circles) and the residual error (colorful circles) are shown. 
The individual metric $p_{\rm ML}$ are calculated at each orbit prediction result, and then used to color the residual errors in the figures. 
\Cref{subfig:reduced} shows all the data points whose errors are reduced by the trained SVM model, and \cref{subfig:increased} shows all the data points on which the ML-modification fails, i.e., the residual errors are increased compared with the true errors, with all the $p_{\rm ML}$ larger than 500\% is colored by yellow. 
Comparing the two figures reveals that the trained SVM model works well on the majority of the testing data point. 
In \cref{subfig:increased}, both the true error and the residual error are relatively small compared with those in \cref{subfig:reduced}, and most failures of the ML approach occur when $\Delta t \leq 1$ days. 
We expect that is due to the fact the orbit prediction is accurate enough and below the modification capability of the trained SVM model. 
However, further studies are required to draw concrete conclusions. 

\begin{figure}[!htbp]
	\centering
	\includegraphics[width=0.7\textwidth]{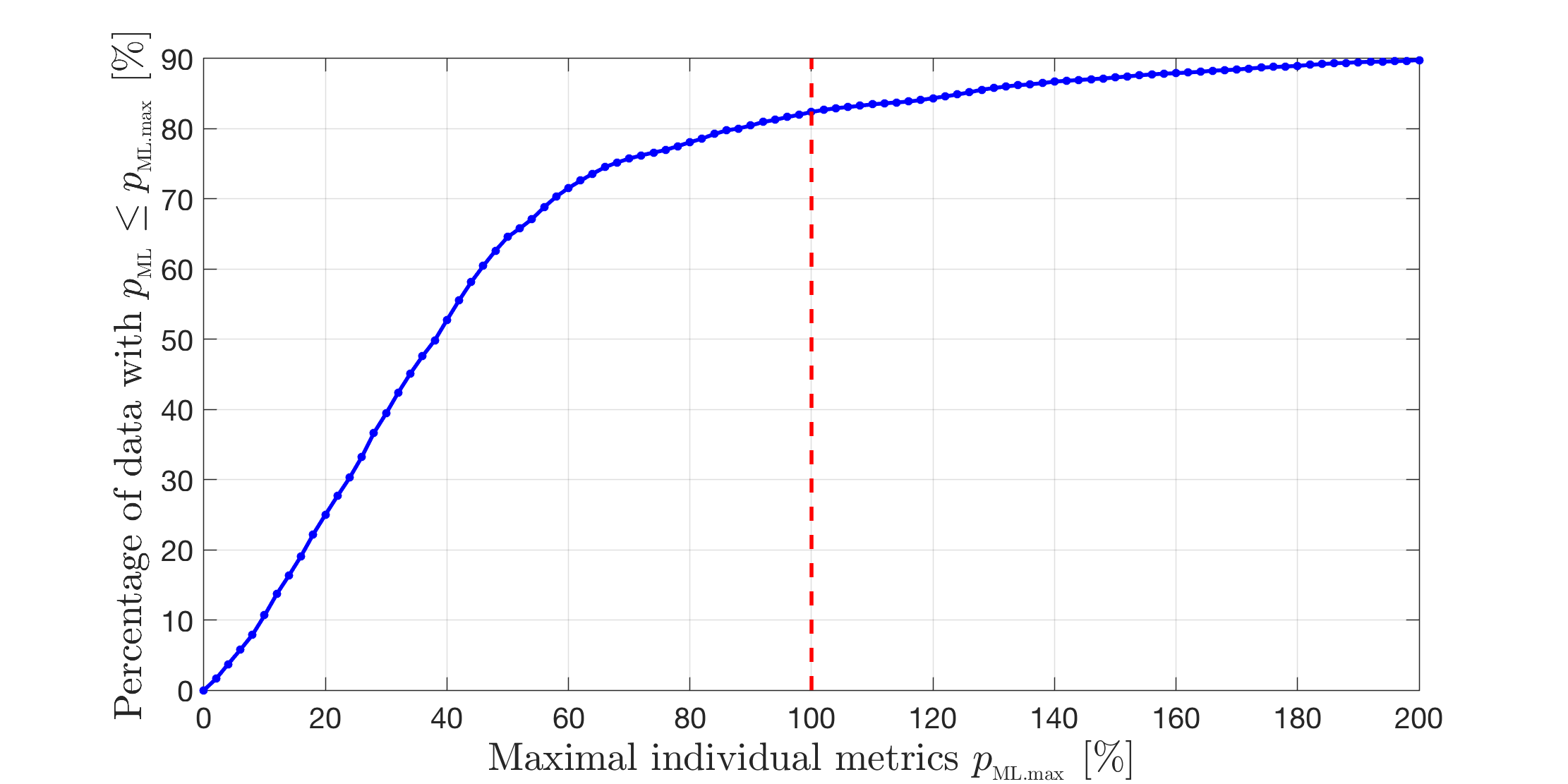}
	\caption{Distribution of the individual metric $p_{\rm ML}$.}
	\label{fig:LEO1 pML densitiy}
\end{figure}

\Cref{fig:LEO1 pML densitiy} demonstrates the distribution of the individual metric $p_{\rm ML}$. 
The vertical axis is the percentage of the data point in the whole training data, whose $p_{\rm ML}$ is less than the corresponding $p_{\rm ML.max}$ represented by the horizontal axis. 
For example, the intersection of the curve and the dashed line indicates that for about 80\% percent of the testing data, the percentage of the residual error is less than 100\%, meaning that their prediction errors have been successfully reduced by the trained SVM model. 
So, generally, the ML approach can improve the orbit prediction accuracy at the majority of the future epochs. 
Since the results of the individual metric $p_{\rm ML}$ on different RSOs are similar, for brevity, the discussions on other RSOs will be omitted in the following paper. 

%Then the residual error is depicted according to p_ML. 

%The individual metric p_ML will be 0 if the orbit prediction error is completely compensated.

\FloatBarrier
\subsection{Results on Other RSOs}

To further demonstrate the capability of the trained ML model, six other RSOs with different orbits are studied. 
These RSOs are chosen from the International Laser Ranging Service (ILRS)~\citep{pearlman_international_2002}, and the parameters are extracted from the TLE catalog, as summarized in \cref{tab:varied_RSO_parameters}. 
The orbit altitudes increase from left to right in \cref{tab:varied_RSO_parameters}, and the LAGEOS-1 has the highest altitude of about 5860 km. 
The simulated RSOs are based on the two-line element (TLE) data of the real satellites, and are assumed with the same area-to-mass ratio of 0.05 m\textsuperscript{2}/kg, drag coefficient of 2.2, and reflection coefficient of 1.25, which are identical to those in \cref{tab:RSO_parameters}. 
For simplicity, we will refer to these RSOs by the name of their base RSOs. 
However, we note that these simulated RSOs do not reflect true orbit information of the base RSOs, because the parameters in \cref{tab:varied_RSO_parameters} are only used as the initial conditions for simulations in this paper. 

\begin{table*}[htb]
	\centering
	\caption{Parameters of simulated RSOs, based on orbits extracted from TLE data retrieved on April 14, 2017.}
	\label{tab:varied_RSO_parameters}
	%	\makebox[\textwidth][c]{
	\footnotesize
	\begin{tabular}{lllllll}
		\toprule
		Base RSO                           & SWARM-A & LARETS & STELLA   & HAIYANG 2A & LARES  & LAGEOS-1 \\ \midrule
		NORAD catalog number               & 39452   & 27944  & 22824    & 37781      & 38077  & 8820     \\
		Orbit altitude [km]                & 460     & 691    & 804--812 & 971        & 1450   & 5860     \\
		Semi-major axis $a$ [km]           & 6820.0  & 7060.2 & 7167.5   & 7340.6     & 7813.9 & 12268.9  \\
		Eccentricity $e$                   & 0.00144 & 0.0002 & 0.0010   & 0.0012     & 0.0001 & 0.0045   \\
		Inclination $i$ [deg]              & 87.36   & 98.204 & 98.6     & 99.35      & 69.5   & 109.84   \\
		Argument of perigee $\omega$ [deg] & 91.51   & 81.13  & 284.18   & 65.83      & 133.44 & 278.75   \\
		RAAN $\Omega$ [deg]                & 151.01  & 355.07 & 177.33   & 230.41     & 58.02  & 151.20   \\
		Mean anomaly $\nu$ [deg]           & 297.75  & 59.94  & 176.41   & 169.42     & 151.57 & 313.31   \\
		%			Area-to-mass ratio             & 0.05   & 0.05   & 0.05     & 0.05       & 0.05   & 0.05     \\
		%			Drag coefficient $C_d$         & 2.2     & 2.2    & 2.2      & 2.2        & 2.2    & 2.2      \\
		%			Reflection coefficient $C_r$   & 1.25    & 1.25   & 1.25     & 1.25       & 1.25   & 1.25     \\ 
		\bottomrule
	\end{tabular}
	%	}
\end{table*}

\begin{figure}[!htb]
	\centering
	\newcommand{\tempOne}[2]{\subfigure[#1]{\includegraphics[trim={22cm 0 0 0}, clip=true, width=0.45\linewidth]{#2} }}
	\tempOne{SWARM-A.}{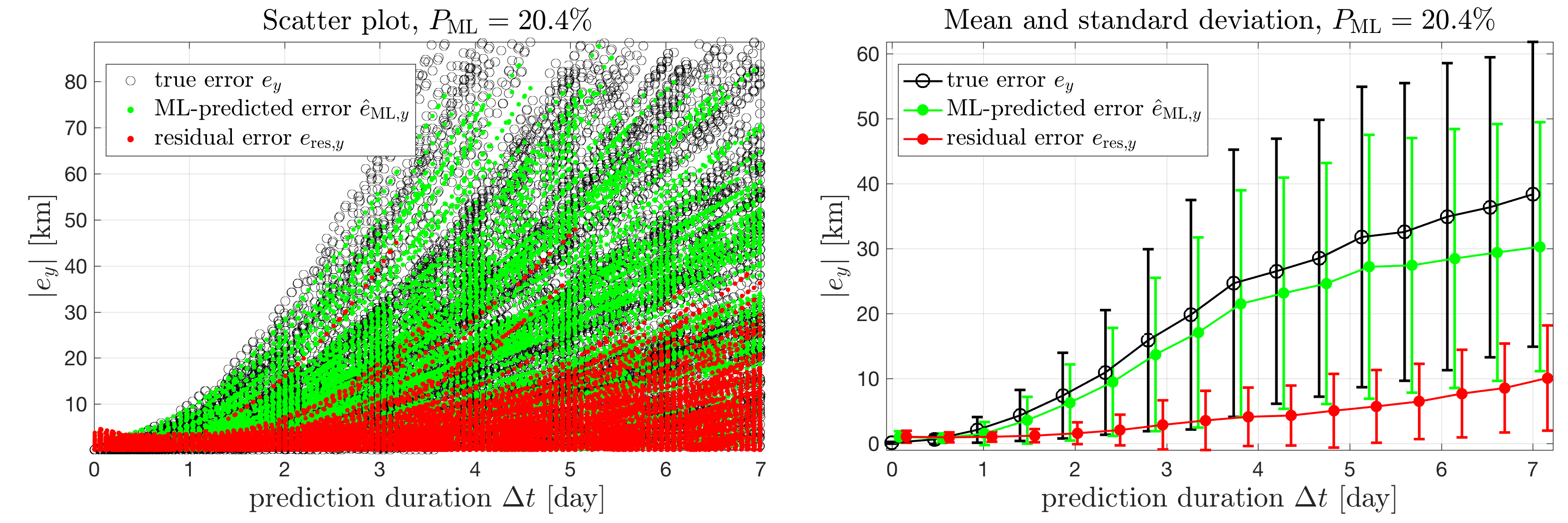}
	\tempOne{LARETS.}{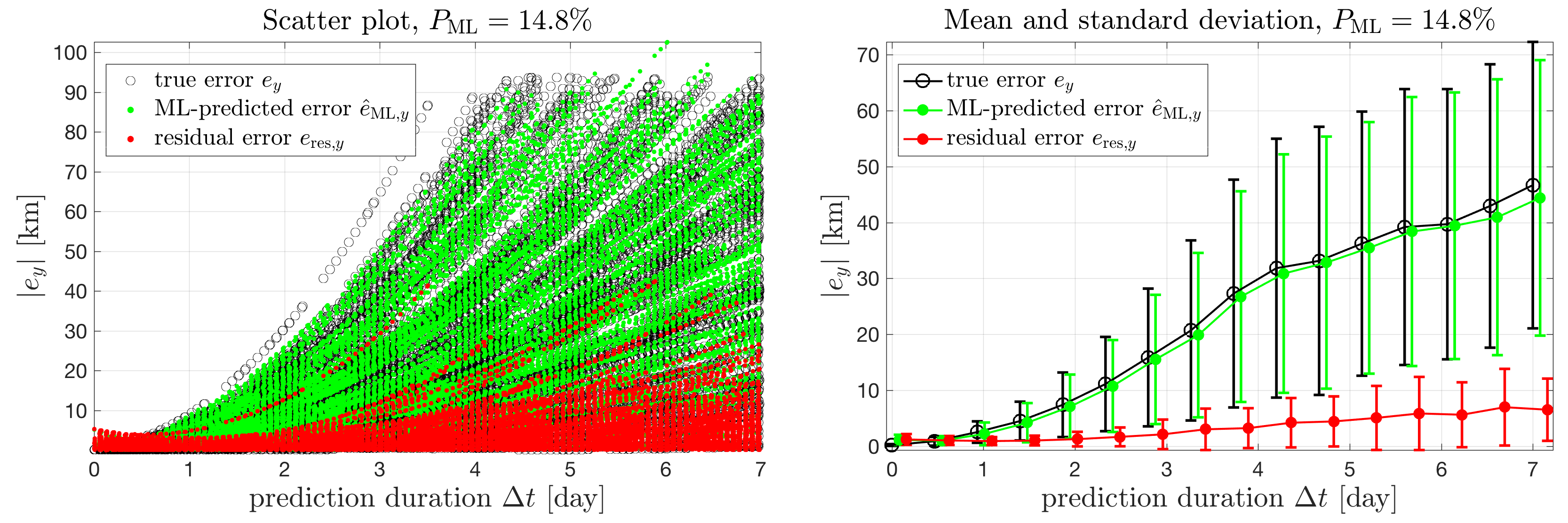} \\
	\tempOne{STELLA.}{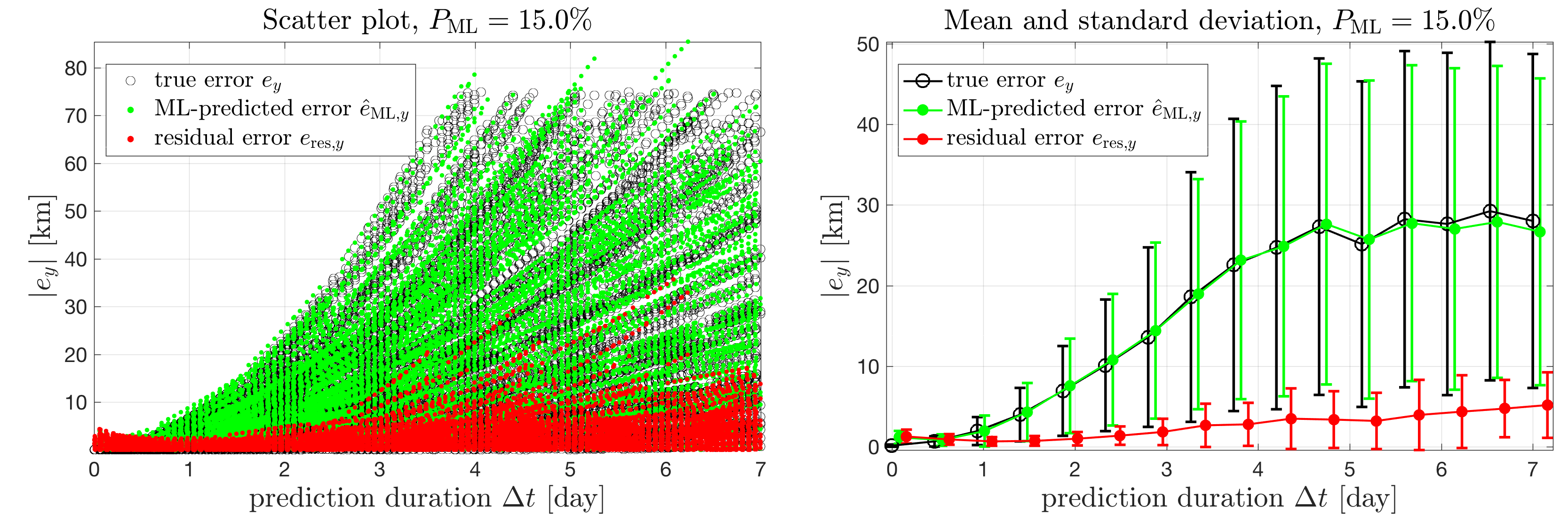}
	\tempOne{HAIYANG 2A.}{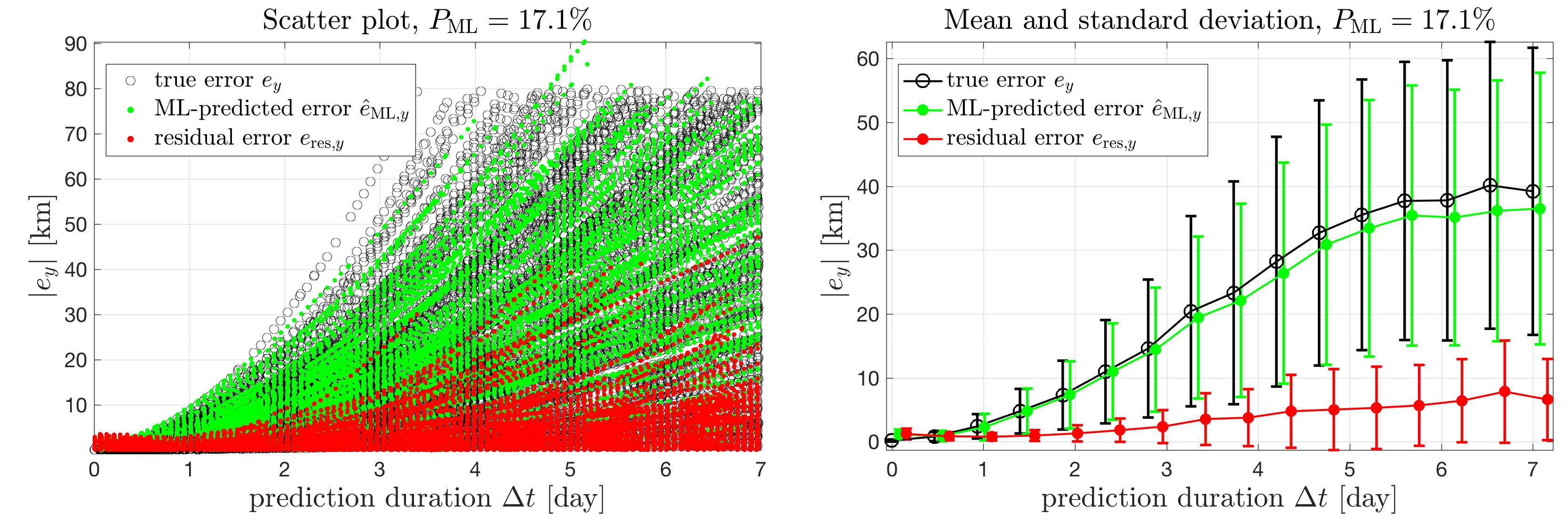} \\
	\tempOne{LARES.}{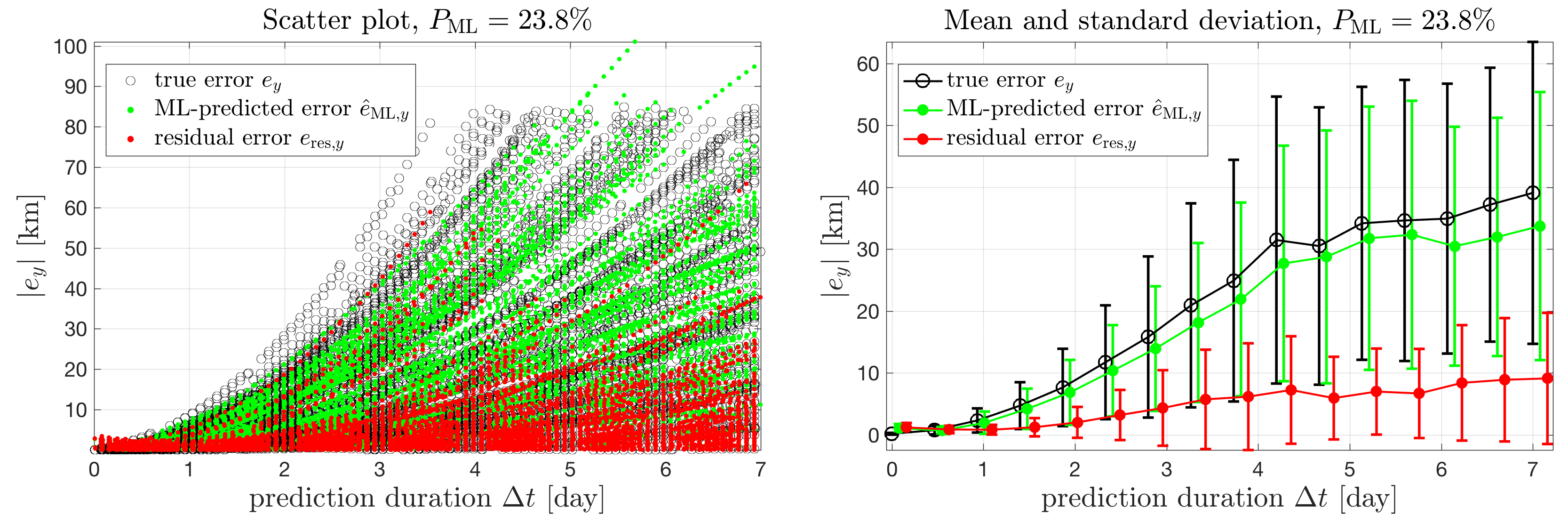}
	\tempOne{LAGEOS-1.}{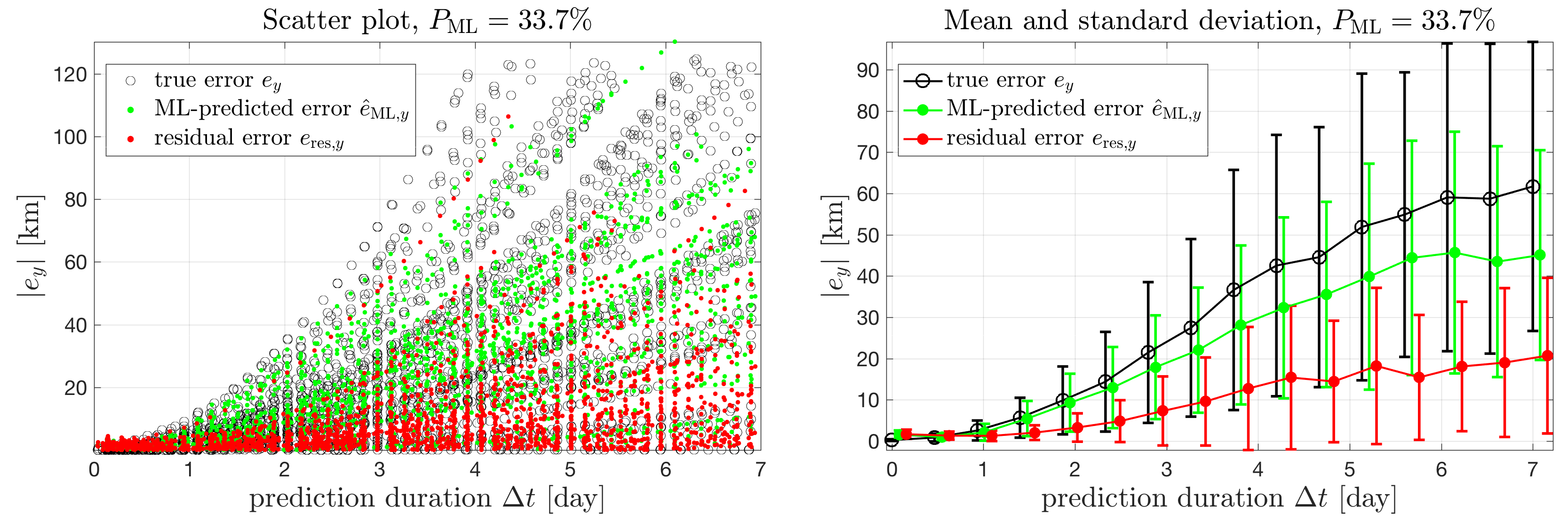}
	\caption{Results of the ML approach applied on RSOs with different orbit altitude, when generalizing to future epochs.}
	\label{fig:varied RSOs}
\end{figure}

The results of generalization capability to future epochs of these six new RSOs are shown in \cref{fig:varied RSOs}. 
For all the RSOs, both the mean values and the standard deviations have been greatly reduced by the trained SVM models. 
The performance of the ML approach varies with the RSO's orbit, and further studies are necessary to reveal the detailed relationship.

\section{Generalization to Nearby RSOs}
\label{sec:generalize to other RSOs}

In this section, we investigate another generalization capability that is of potential interest for the SSA applications. 
In practice, we may only have accurate historical data about a limited number of RSOs. 
And the question is whether an error model learned from such RSOs can be applied to improve orbit prediction accuracy of other RSOs, for which not enough data is available to build the ML model. 

\subsection{Varying Orbit Inclination}

Here we vary the orbit of the training RSO (based on the ISS in \cref{tab:RSO_parameters}) to evaluate the generalization capability of the trained SVM model to nearby RSOs. 
The inclination of the learned RSO is $51.6393^\circ$. 
We have changed the inclination to $48^\circ$, $50^\circ$, $52^\circ$ and $54^\circ$ respectively, but have kept all the other orbital elements unchanged. 
All the new RSOs are propagated from the same starting epoch, and then are measured, estimated, and predicted within 0--50 days. 
The training data is the error data collected in the first 30 days of the training RSO. 
And then the trained model is used to modify the orbit prediction of the other RSOs in the following 10 and 20 days. 
In other words, the testing data of the trained SVM model is the error collected in the 30--40 days and 30--50 days of the other RSOs. 

\begin{figure}[htbp]
	\centering
	\includegraphics[width=0.9\linewidth]{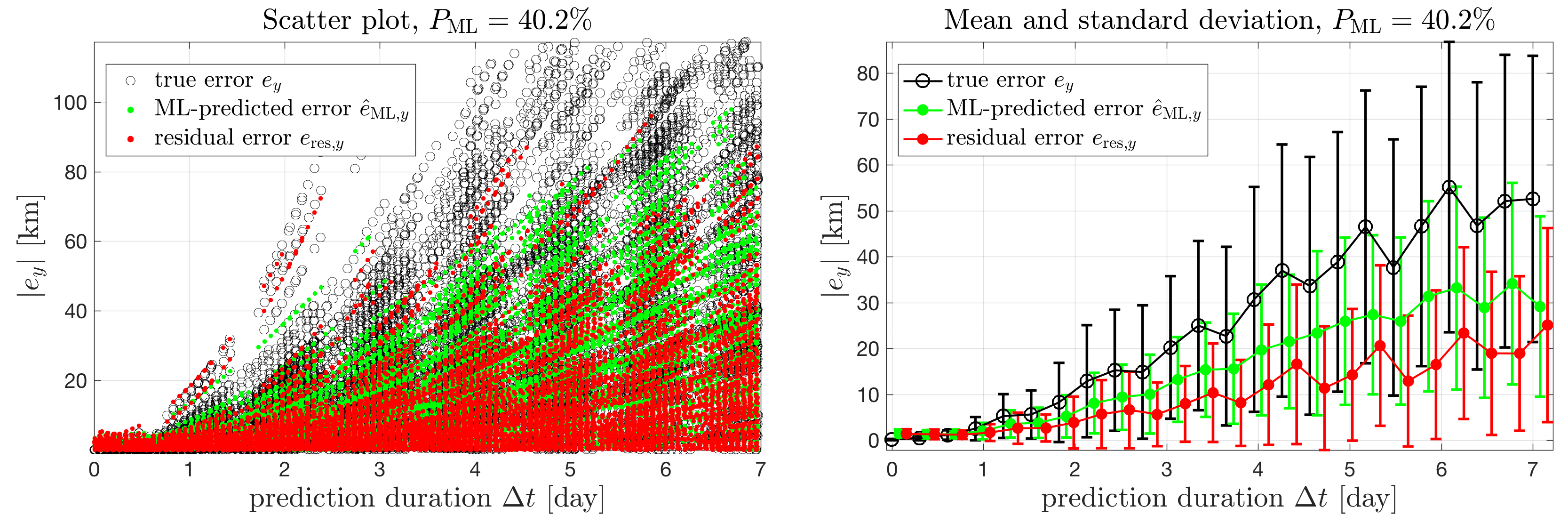}
	\caption{Results of the trained SVM model on a new RSO with $i=50^{\circ}$.}
	\label{fig:leo1to3-ey-day0-30-30-40-testing-absolute}
\end{figure}

The results of modifying the prediction error of the new RSO with $i=50^{\circ}$ in following 10 days are shown in \cref{fig:leo1to3-ey-day0-30-30-40-testing-absolute}, where the testing data is the error data in 30--40 days. 
The mean error and the standard deviation have been greatly reduced. 
Compared with the performance on the training RSO, shown in \cref{fig:along track error}, the performance metric $P_{\rm ML}$ has increased from 35.5\% to 40.2\%, which indicates the improvement still exists but the performance becomes worse. 

\begin{figure}[htbp]
	\centering
	\includegraphics[width=0.8\linewidth]{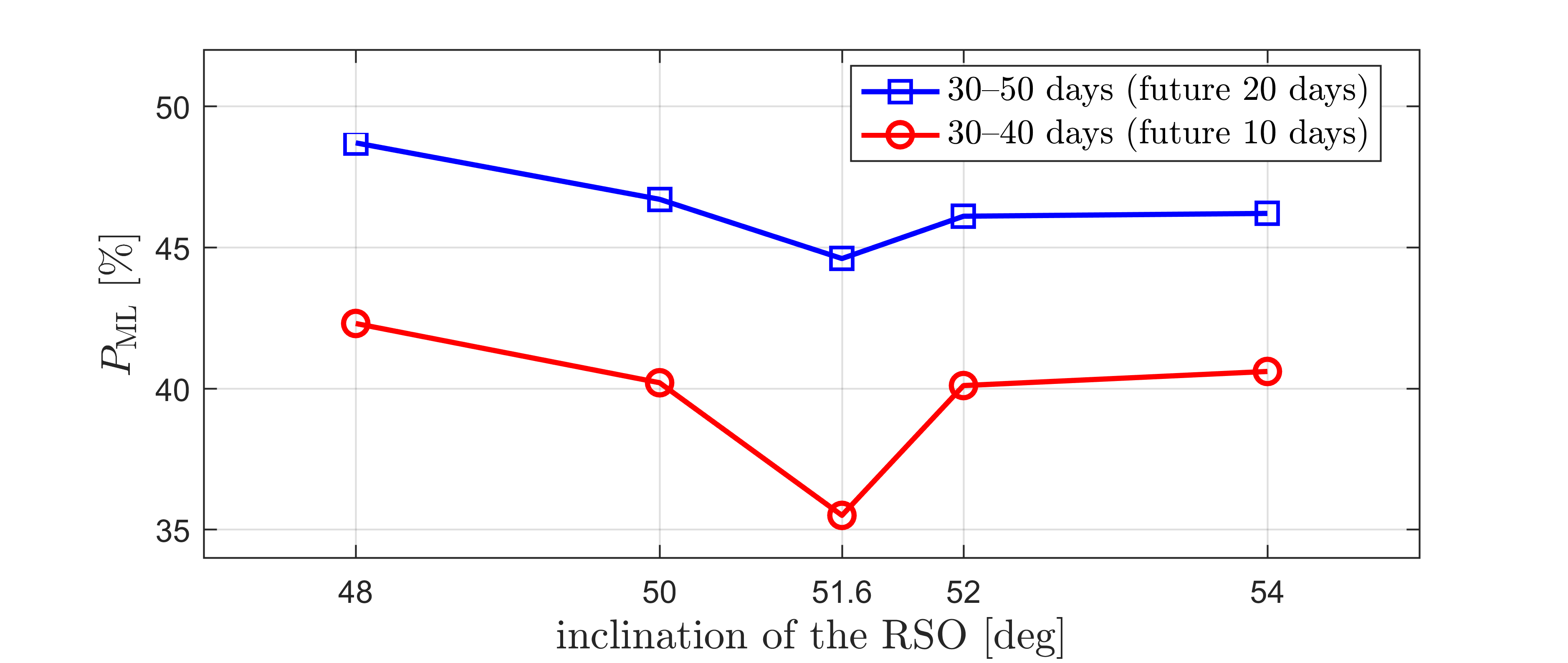}
	\caption{ML-modified results on nearby new RSOs with $i$ ranging from $48^\circ$ to $54^\circ$.}
	\label{fig:LEO_01 modify LEO_03--6}
\end{figure}

The results for all the five RSOs are summarized in \cref{fig:LEO_01 modify LEO_03--6}. 
The performance metric $P_{\rm ML}$ on the training RSO (with $i=51.6393^\circ$) is the smallest. 
Compared with the training RSO, $P_{\rm ML}$ on other new RSOs are larger. 
But even in the worse case, $P_{\rm ML}$ is still less than 50\%, which means more than half of the prediction errors can be corrected by the trained SVM model. 
Furthermore, the results show that the larger the deviation of the inclination from the RSO that has been learned, the worst the performance. 
Again, such results are intuitive, but nice to be seen from the machine learning results.

%\FloatBarrier
\subsection{Varying Orbit Altitude}

Another important parameter for RSOs in LEO is the orbit altitude. 
Here, the semi-major axis of the simulated RSO based on ISS is increased. 
The orbit prediction results in day 30--40 are used as the testing data, and the SVM model is still trained using the training data of the original RSO in day 0--30. 

\begin{figure}[htbp]
	\centering
	\includegraphics[width=0.9\linewidth]{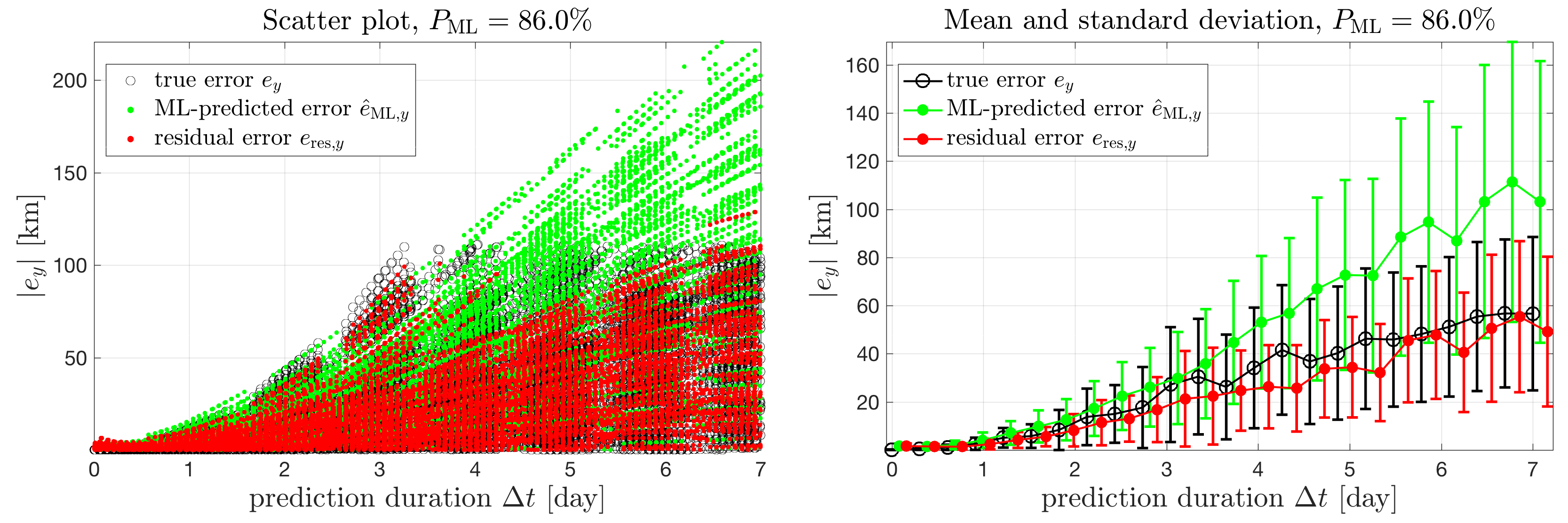}
	\caption{Results of the trained SVM model applied to a new RSO with $a=6833.34$ km (with an 50 km increment on the original semi-major axis in \cref{tab:RSO_parameters}).}
	\label{fig:varying semi-major axis}
\end{figure}

The results of the trained SVM on a new RSO with $a=6833.34$ km are shown in \cref{fig:varying semi-major axis}. 
The semi-major axis of the new RSO is increased by 50 km, and the metric $P_{\rm ML}$ is 86.0\%, which means the orbit prediction errors are only reduced by about 14.0\%. 
The performance of the trained SVM model is not as good as that in \cref{fig:along track error}, but it is expected because the simulated RSO is in LEO and an increment of 50 km on the semi-major axis can introduce a lot of differences. 

\begin{figure}[htbp]
	\centering
	\includegraphics[width=0.75\linewidth]{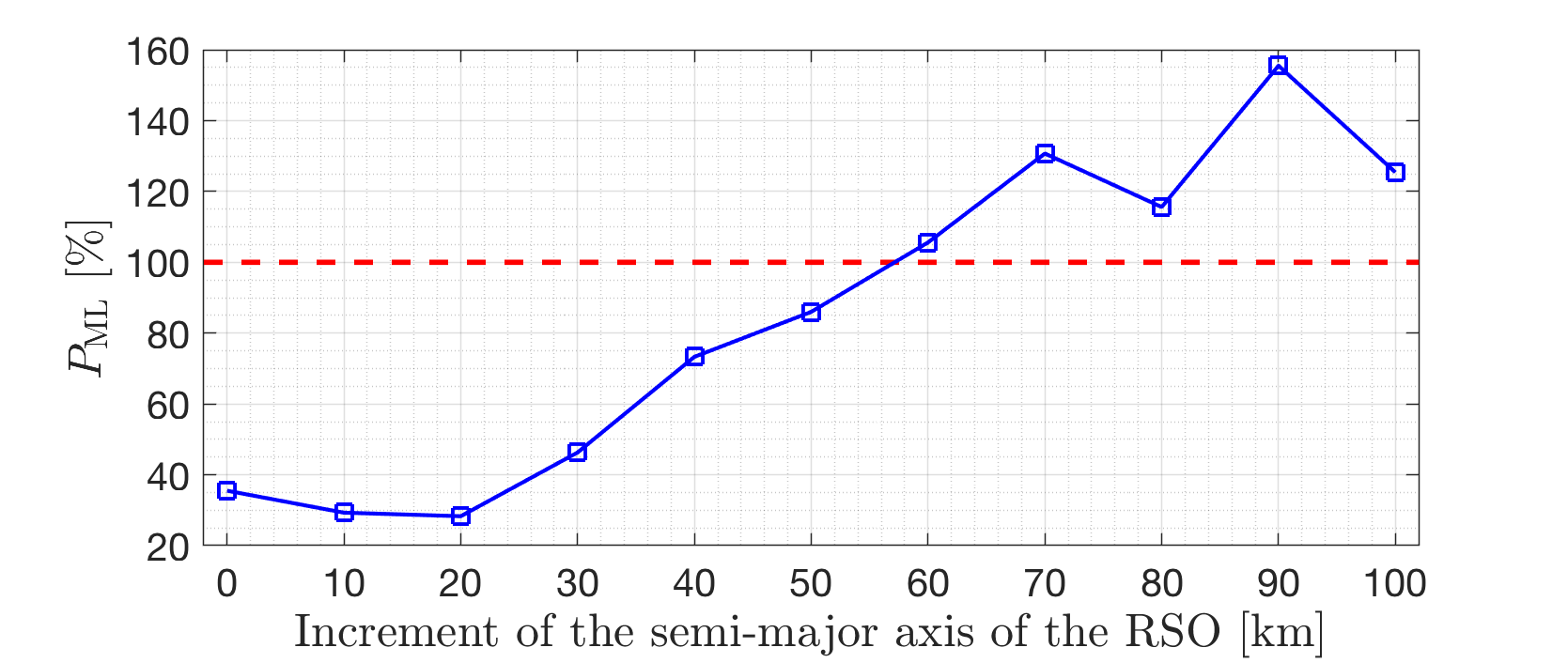}
	\caption{ML-modified results on nearby new RSOs with an increment of $a$ ranging from 0 to 100 km.}
	\label{fig:LEO1 modify LEO7-12}
\end{figure}

The performance metric $P_{\rm ML}$ of RSOs with other increments are demonstrated in \cref{fig:LEO1 modify LEO7-12}. 
We remark that a decrease of the semi-major axis of the original RSO will result in a reentry in less than 40 days, so only the increment of semi-major axis is considered. 
It is observed that $P_{\rm ML}$ is decreasing at the beginning. 
This could be caused by many factors, such as the randomness of the dataset, and the possible variation of the atmosphere density in this region. 
Meanwhile, the $P_{\rm ML}$ increases quickly as the increment of semi-major axis becomes larger than 30 km. 
There are also some fluctuation on the performance when the increment is larger than 60 km. 
However, for these cases the metric are all larger than 100\%, meaning that the ML approach does not improve the orbit prediction accuracy. 
For the moment, we can only conclude that the trained SVM model can be generalized to other RSOs with different semi-major axes, but it will fail when the variation is too large.

The results in this section are interesting as there is no information directly related to the new RSOs in the training dataset, but the ML approach could still be used to improve the orbit prediction accuracy. 
%
%The learning variables in the designed data structure captures much information about the system. 
Some knowledge learned for the data depends on RSO's properties such as the orbit altitude and inclination. But some other knowledge can be universal to other RSOs, such as the errors of the assumed model, measurement, and estimator, which depend on the environment and the catalog system. 
Our conjecture is that the proposed ML approach does not distinguish these differences, so the trained SVM model captures not only the information about the specific RSO that has been used for training, but also some information of the system. 
%Further studies will examine these details about the ML approach. 

\section{Conclusions}

In this paper, a new approach of using the supervised machine learning (ML) method to improve the orbit prediction accuracy of the resident space objects (RSOs) is proposed. 
The framework of incorporating an ML model in the orbit prediction process is formulated and the structure of the dataset is designed based on analyzing the prediction errors and also other possibly available parameters. 

A simulation-based space catalog environment is developed to evaluate the performance of the proposed approach, and the support vector machine (SVM) model is chosen as the machine learning algorithm in the paper. 
RSOs in low Earth orbit are simulated to examine the performance of the SVM model. 
The results demonstrate that the trained SVM model can: 1) improve the information about the same RSO that shares the same time duration as the training data but are not available during training; 
2) improve the prediction accuracy of the same RSO at future epochs;
and 3) improve the prediction accuracy of other nearby RSOs unknown to the trained SVM model. 
It is shown that the trained SVM can reduce more than a half of the prediction error for all the three types of improvement. 
We note that this is an exploration study and further research is required before practical applications of the ML approach. 

Further research is suggested to apply the established framework to real operational data, where the RSOs with large amount of measurement data can be used as the training objects and the trained SVM model can be used to improve orbit accuracy of the other RSOs that share some similar orbit characteristics.

\section*{Acknowledgment}
The authors would acknowledge the research support from the Air Force Office of Scientific Research (AFOSR) FA9550-16-1-0184 and the Office of Naval Research (ONR) N00014-16-1-2729.  
Large-scale simulations have been carried out on the School of Engineering (SOE) HPC cluster at Rutgers University. 
We also appreciate anonymous reviewers for insightful discussions and comments on the paper.

\section*{References}
\bibliographystyle{model4-names}
\bibliography{thislib}

\end{document}